\documentclass[sn-mathphys,Numbered]{sn-jnl}% Math and Physical Sciences Reference Style
\usepackage{graphicx}%
\usepackage{multirow}%
\usepackage{amsmath,amssymb,amsfonts}%
\usepackage{amsthm}%
\usepackage{mathrsfs}%
\usepackage[title]{appendix}%
\usepackage{xcolor}%
\usepackage{textcomp}%
\usepackage{manyfoot}%
\usepackage{booktabs}%
\usepackage{algorithm}%
\usepackage{algorithmicx}%
\usepackage{algpseudocode}%
\usepackage{listings}%
\usepackage[inline, shortlabels]{enumitem}
\usepackage{subcaption}
\usepackage{array}
\usepackage{booktabs}
\usepackage{multirow}
\usepackage{svg}
\usepackage{parskip}
\usepackage{lineno}
\usepackage{setspace}

\begin{document}

\title[Article Title]{Automated Annotation of Scientific Texts for ML-based Keyphrase Extraction and Validation}


\author*[1]{\fnm{Oluwamayowa O.} \sur{Amusat}}\email{ooamusat@lbl.gov}

\author[2]{\fnm{Harshad} \sur{Hegde}}\email{hhegde@lbl.gov}

\author[2]{\fnm{Christopher J.} \sur{Mungall}}\email{cjmungall@lbl.gov}

\author[1]{\fnm{Anna} \sur{Giannakou}}\email{agiannakou@lbl.gov}

\author[3]{\fnm{Neil P.} \sur{Byers}}\email{npbyers@lbl.gov}

\author[1]{\fnm{Dan} \sur{Gunter}}\email{dkgunter@lbl.gov}

\author[3]{\fnm{Kjiersten} \sur{Fagnan}}\email{kmfagnan@lbl.gov}

\author[1]{\fnm{Lavanya} \sur{Ramakrishnan}}\email{lramakrishnan@lbl.gov}

\affil*[1]{\orgdiv{Scientific Data Division}, \orgname{Lawrence Berkeley National Laboratory}, \orgaddress{\city{Berkeley}, \postcode{94720}, \state{California}, \country{USA}}}

\affil[2]{\orgdiv{Division of Environmental Genomics and Systems Biology}, \orgname{Lawrence Berkeley National Laboratory}, \orgaddress{\city{Berkeley}, \postcode{94720}, \state{California}, \country{USA}}}

\affil[3]{\orgdiv{DOE Joint Genome Institute}, \orgname{Lawrence Berkeley National Laboratory}, \orgaddress{\city{Berkeley}, \postcode{94720}, \state{California}, \country{USA}}}

\abstract{Advanced omics technologies and facilities generate a wealth of valuable data daily; however, the data often lacks the essential metadata required for researchers to find and search them effectively. The lack of  metadata poses a significant challenge in the utilization of these datasets. Machine learning-based metadata extraction techniques have emerged as a potentially viable approach to automatically annotating scientific datasets with the metadata necessary for enabling effective search. Text labeling, usually performed manually, plays a crucial role in validating machine-extracted metadata. However, manual labeling is time-consuming and not always feasible; thus, there is an need to develop automated text labeling techniques in order to accelerate the process of scientific innovation. This need is particularly urgent in fields such as environmental genomics and microbiome science, which have historically received less attention in terms of metadata curation and creation of gold-standard text mining datasets.

In this paper, we present two novel automated text labeling approaches for the validation of ML-generated metadata for unlabeled texts, with specific applications in environmental genomics. Our techniques show the potential of two new ways to leverage existing information about the unlabeled texts and the scientific domain. The first technique exploits relationships between different types of data sources related to the same research study, such as publications and proposals.  The second technique takes advantage of domain-specific controlled vocabularies or ontologies. In this paper, we detail applying these approaches in the context of environmental genomics research for ML-generated metadata validation. Our results show that the proposed label assignment approaches can generate both generic and highly-specific text labels for the unlabeled texts, with up to 44\% of the labels matching with those suggested by a ML keyword extraction algorithm.}




\maketitle

\section{Introduction}\label{intro}

\begin{figure*}
    \centering
    \includegraphics[scale=0.7]{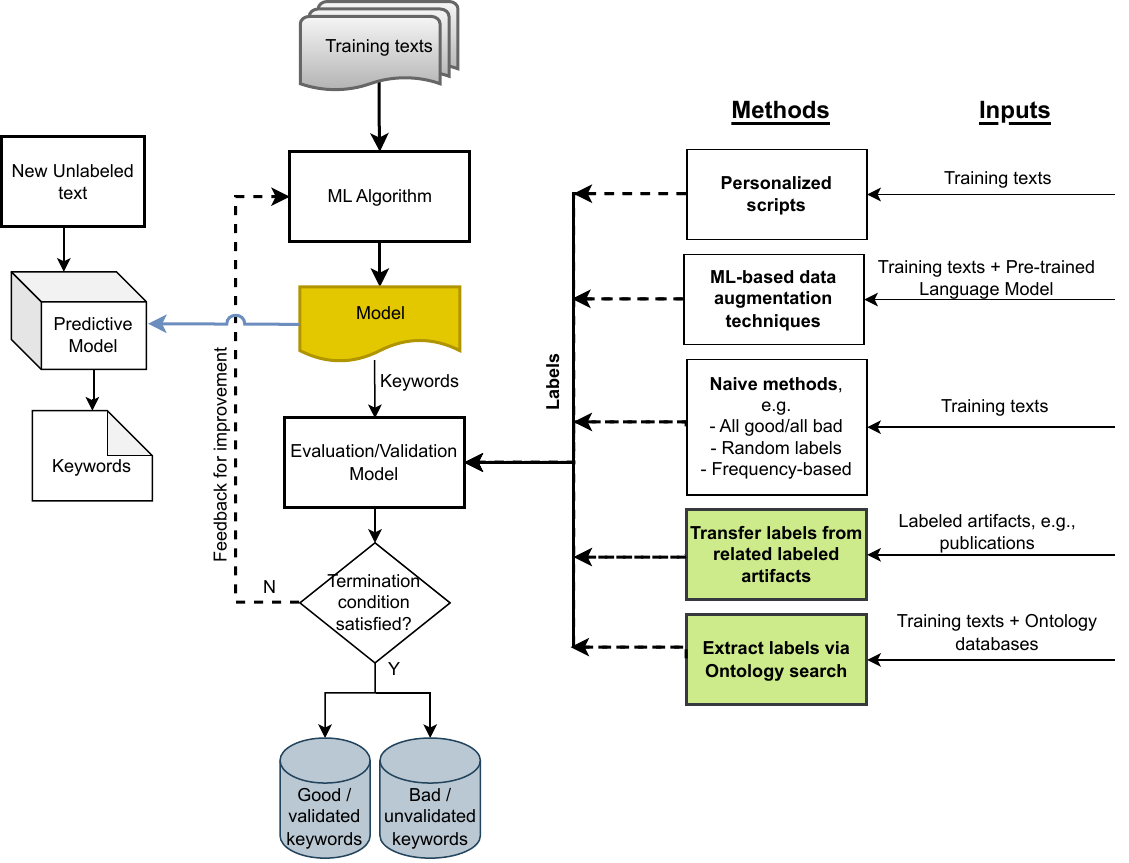}
    \caption{Potential approaches for validating ML-generated keywords for unlabeled texts when human labels are unavailable. Our proposed approaches are shown in green.}
    \label{fig:labeling-unsupervised-ml}
\end{figure*}

High throughput omics technologies such as genome sequencing produce a wealth of data in domains ranging from human health, biosurveillance, and environmental microbial ecology. However, these data frequently lack the metadata necessary for scientists to search, integrate, and interpret this data appropriately. Manual labeling is time-consuming and error prone, and is made more difficult by the paucity of gold-standard training sets, particularly in domains such as environmental science. Machine learning (ML)-based metadata extraction techniques have emerged as a potential solution to the challenge of automatically annotating scientific artifacts such as documents and images with the metadata necessary for enabling effective search \cite{Gunther2018, Papagiannopoulou2019}. However, these ML-based approaches depend on the existence of text labels for either training the models (supervised methods) or validating the ML-generated metadata (unsupervised methods). Thus, applying these approaches to unlabeled scientific texts requires that text labels be assigned to the documents before the quality of any ML-generated keywords can be determined (Figure \ref{fig:labeling-unsupervised-ml}). This can be challenging since text labeling is still a largely manual process; it is tedious, time-consuming, often costly, and may not always be feasible \cite{Ding2022}. The biomedical community has produced a number of high-quality annotation corpora such as CRAFT \cite{Bada2012}, but these do not cover many relevant concepts in environmental genomics, ranging from sequencing methodologies through to microbial taxa and environmental concepts such as global biogeochemical fluxes, environmental contamination, and so on.

Automation of the text labeling process is necessary; developing automated text labeling techniques will be crucial in accelerating scientific innovation. 
Figure \ref{fig:labeling-unsupervised-ml} presents some potential approaches for automated validation of ML-generated keywords. Naïve techniques which validate text labels based solely on heuristic or statistical rules are easy and quick to implement and only require the training texts as inputs. However, these techniques create no baseline labels themselves but select from the limited set provided by the ML algorithm, thereby implicitly incorporating assumptions about the quality of the ML-generated keywords. Thus, they generally provide poor assessments of keyword quality. Personalized scripts may work well in individual cases; however, such localized scripts tend to be non-transferable and often are not extensible to similar cases or problems. ML-based data augmentation techniques based on pre-trained language models have shown much promise for text labeling and classification \cite{Kumar2020, Papanikolaou2020}; however, they are computationally expensive and data-intensive \cite{Lee2021}, and do not always account for the domain-specific nature of scientific texts.
%While there is some research on automated text augmentation, the focus has primarily been on the use of language models which are computationally intensive, do not always account for the domain-specific nature of scientific texts, and often fail to cover the full diversity and complexity of real examples \cite{Lee2021}.

Our ideas about automating text labeling are inspired by the nature of scientific texts, data, and search at research and user facilities. Scientific artifacts are often related to one another; for example, research that eventually appear in publications often start out as research proposals, technical reports or dissertations. Leveraging the links between these existing artifacts provide a potentially rich source of information. Also,  keywords and search terms for scientific artifacts are typically descriptive terms from a controlled vocabulary or language - domain-specific words or phrases that are generally accepted within the research field of interest. For most scientific domains, these controlled vocabularies may be found in Ontologies. Ontologies are systems of carefully defined terminologies that provide information about how entities or concepts within a subject domain are related \cite{brank2005}, making them a good source of domain-specific knowledge. 

In this paper, we propose two semi-automated techniques for generating text labels to validate ML keywords in order to enhance search. The first approach, called \textit{artifact linking},  is based on exploiting the relationships between different types of data artifacts related to the same research. Establishing direct relationships between artifacts is extremely powerful since it enables the transfer of text labels between the related labeled and unlabeled artifacts, making it a rich source of metadata.  The second approach explores the concept of label assignment to the unlabeled texts from controlled domain-specific vocabularies such as ontologies. The technique exploits the fact that the most relevant text labels for scientific texts will contain domain-specific language that will be present in the ontologies relevant to that domain.  The domain-specific relevance of the words and phrases in the unlabeled texts are determined via Ontologies, with a frequency-based approach used to assign keyphrases to the texts.  

In this paper, we present novel non-human, non-ML techniques for validating ML-generated metadata for unlabeled texts such as proposals (narrative descriptions of proposed hypothesis-driven research). Additionally, we compile parts of existing vocabularies and ontologies, primarily from the Open Bio Ontologies (OBO) Library \cite{Jackson2021} into a novel application ontology called BERO (Biological and Environmental Resource Ontology), intended for applications such as environmental genomics.

Building on the foundations and infrastructure already in ScienceSearch \cite{Rodrigo2018}, the approaches are applied to artifacts from the Joint Genome Institute (JGI), a DOE user facility that provides integrated high-throughput sequencing, DNA design and synthesis, metabolomics, and computational analysis for advancing genomics research, with an emphasis on elucidating environmental systems, and the roles of plants, microbes, in these environmental systems. JGI provides scientific researchers with access to genomic sequencing capabilities and equipment available at a select few research centers in the world and as a result, produces many labeled and unlabeled data artifacts. Developing semi-automated text labeling techniques for JGI's unlabeled data artifacts will enhance its search and indexing capabilities, thus potentially accelerating scientific discovery in genomics research. While the techniques presented are being demonstrated in the specific context of genomics, the techniques are general and can be applied to other scientific domains. Our semi-automated techniques for text labeling advance the state-of-the-art in keyphrase extraction in two ways.
\begin{itemize}
\item We develop computationally inexpensive,  non-human, non-machine labeling approaches for the validation of ML-generated keywords for unlabeled texts that removes the bottleneck of absent labels/keywords, one of the primary challenges associated with extracting relevant keywords with high accuracy \cite{Campos2020}. 
\item Our approaches primarily exploit real available public (human)
knowledge such as related scientific works and controlled
vocabularies. This has several advantages, including being particularly suited to handling and exploiting the
domain-specific nature of scientific texts.
% \item the development of potential non-human labeling techniques is a further step in the direction of full automation of the keyphrase extraction process. 
\end{itemize}

The rest of the paper is organized as follows: Section \ref{rel-work} provides important background information. Section \ref{methodology} presents
our framework for automated label generation and details our two proposed approaches for automated label generation. It also presents the ML-based keyphrase evaluation approach employed, along with information about some of the other decisions made regarding label ranking and hyperparameter optimization. The results obtained for our proposed approaches are presented and discussed in Section \ref{Results}, and some important observations about the results and methods proposed are discussed in Section \ref{Discussion}. We conclude by presenting a review of related work in Section \ref{RelatedWork}.

\section{Background}\label{rel-work}

Our text labeling approaches have been developed and integrated into the ScienceSearch pipeline for automated metadata generation. In this section, we first present a summary of how the unlabeled JGI data artifacts that require labels are generated and stored. We then present a brief overview of the ScienceSearch infrastructure, SciKey, and how our work fits into the framework.

\subsection{JGI Data Generation and Management}

The sequencing and computational analysis capabilities offered by JGI leads to the generation of massive volumes of labeled and unlabeled data artifacts.

% I'm not really comfortable with the above; yes, JGI generates
% massive volumes of data artefacts, but our approach is only on the proposal
% text itself which is a fraction of this... we're not indexing fastq
% files here -- CJM

The use of JGI's facilities begins with an application process. Applicants submit proposal documents containing details and justifications for the proposed research, with the submitted proposals evaluated by expert domain scientists as to their scientific significance and relevance to DOE science missions. The approved projects send their samples to JGI where the samples are sequenced, with the processed results sent back to the researchers. The sequencing results usually end up being published in some form, either as a stand-alone resources (e.g. technical reports, thesis), or as part of research papers.  

As part of the JGI data management process, approved proposals are organized and stored in JGI's Work Initiation Process (\emph{WIP}) system. \emph{WIP} is JGI's proposal submission and management system that allows the scientific community to request sequencing or synthesis resources from JGI and submit metadata required for proposal submission. Each proposal is assigned a unique integer Proposal ID that links projects and samples associated with the proposal. Examples of fields present for the proposals include author names, affiliations, project descriptions, and approval dates.

%As part of the JGI data management process, the approved proposals are organized and stored in \emph{JAMO} (JGI Archive and Metadata Organizer), an in-house metadata and file repository system  \textbf{[JAMO ref]}. JAMO contains user-supplied data (and where available, metadata) for all files and data artifacts produced within the Joint Genome Institute. A few examples of the fields present in JAMO for proposals are the author names, affiliations, project description, and approval date. 

The proposals stored in \emph{WIP} are the unlabeled scientific texts of interest in this work. The JGI proposals typically lack metadata related to their scientific contents, so developing techniques to automatically generate text labels is critical to facilitating search and indexing.

\subsection{ScienceSearch}
ScienceSearch \cite{Rodrigo2018} is a generalized scientific search infrastructure that uses machine learning (ML) to capture metadata from data and surrounding artifacts. The ScienceSearch platform enhances search capabilities across several scientific domains including genomics, earth sciences, and microscopy. The ScienceSearch infrastructure comprises four primary components: data ingestion, metadata extraction, search engine, and user feedback. The search capabilities provided by the ScienceSearch infrastructure are critical to advancing scientific data exploration, allowing end users to search across different data artifact types (e.g., publications, proposals, file system paths, images), and provide feedback on automatically generated tags.

Automated metadata extraction occurs in \textit{SciKey}, the metadata extraction component of the ScienceSearch infrastructure. 

\subsection{SciKey}

\begin{figure*}
\centering
\includegraphics[width=1.0\textwidth]{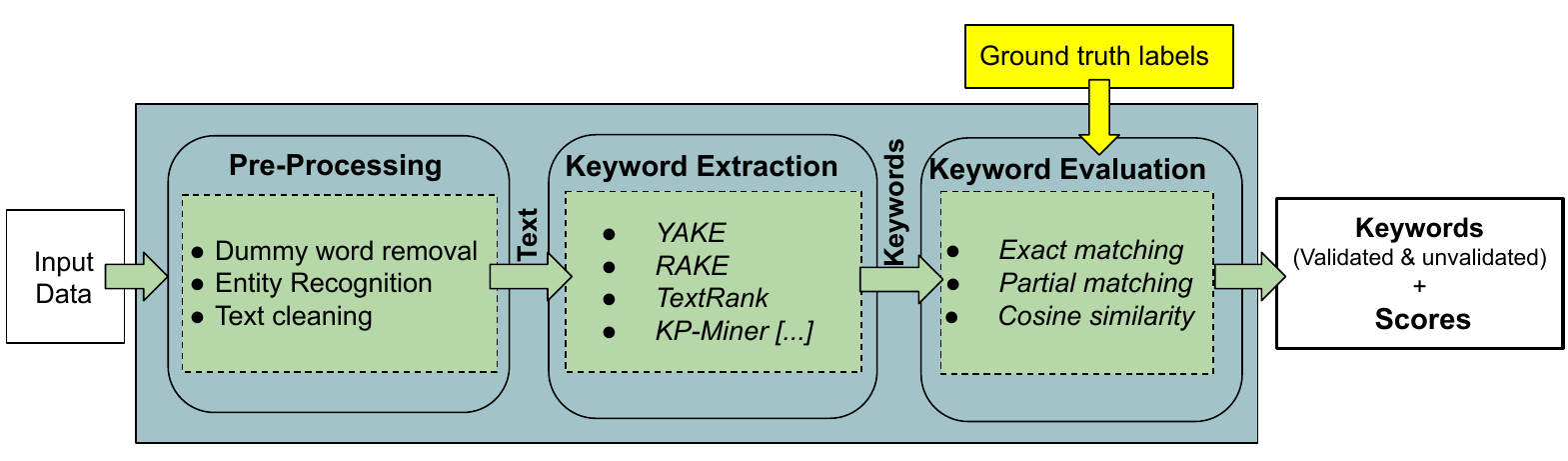}
	\caption{\textit{SciKey's} metadata generation pipeline and sumodules.}
\label{fig:scikey_pipeline} 
\end{figure*}

\textit{SciKey} \cite{Giannakou2024} is a domain-specific, modular, keyword extraction pipeline that incorporates different NLP extraction techniques for automatically generating keywords and keyphrases from scientific datasets. Figure \ref{fig:scikey_pipeline} shows a simplified representation of the three sub-modules that make up \textit{SciKey}: \textit{pre-processing}, \textit{keyword extraction} and \textit{keyword evaluation}. 

The \textit{pre-processing} module prepares the raw input data for NLP ingestion. Scientific texts typically contain domain-specific, non-standard text information such as abbreviations and acronyms, as well as non-text information such as numbers and punctuations. Text sanitization occurs in the text-preparation step: non-text information is removed using pattern-matching approaches such as RegEx, while Named Entity Recognition (NER) techniques and manually-curated lists are employed to handle the non-standard scientific words and concepts. The pre-processing step contains sub-modules for 1) dummy word removal, 2) named entity recognition, and 3) text cleaning.

The \textit{keyword extraction} module generates keywords from the sanitized texts via natural language processing. \textit{SciKey} offers has a suite of unsupervised learning algorithms for NLP keyphrase extraction, including TextRank \cite{mihalcea2004}, RAKE \cite{Rose2010} and YAKE \cite{Campos2020}. The output from this module is a set of unvalidated machine-generated keywords.

The last step for all keyword generation solutions is evaluating the quality of the  ML-generated keywords. The \textit{keyword evaluation} component of the \textit{SciKey} pipeline computes quantitative metrics for the quality of the NLP keywords generated in the \textit{keyword extraction} stage by comparing against a set of provided ground truths (i.e., labels). \textit{SciKey} provides a variety of information retrieval techniques for keyword extraction validation, including exact matching, partial matching, and cosine similarity.

\textit{SciKey} has been demonstrated to work well for labeled datasets (where  ground truth labels are available), with each module in the pipeline allowing for domain-specific customization. However, to take advantage of all the scientific information available, there is a need to extend these capabilities to unlabeled scientific datasets such as proposals and reports (where ground truth labels are not available). The techniques developed in this work are integrated into the \textit{SciKey} pipeline (a key component of the keyphrase extraction step in this work) as a potential solution to this challenge.

\section{Methodology} \label{methodology}

\begin{figure}
	\centering
	\includegraphics[width=1.0\textwidth]{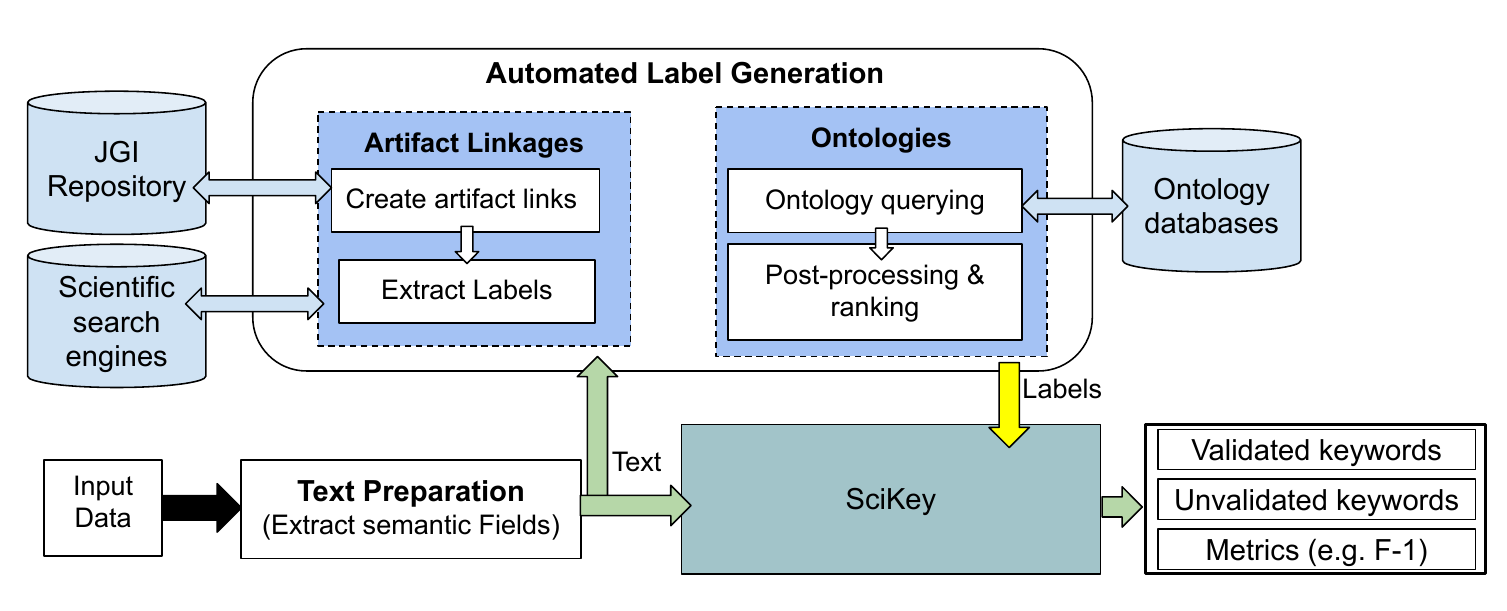}
	\caption{Overview of the automated labeling and metadata generation process. Built on top of the \textit{SciKey} module, the automated label generation process (shown in blue) takes in a text blob containing semantic information and a set of labels that are used for ML-generated keyword validation. Based on these labels, \textit{Scikey} outputs a set of validated ML keywords (i.e "good" labels) and quantitative metrics reflecting keyword quality.}
\label{fig:pfd-autolabels} 
\end{figure}

Figure~\ref{fig:pfd-autolabels} shows our automated label generation process and how it interacts with the \textit{SciKey} pipeline.

The \textit{text preparation} method extracts the semantic information from the raw input data and converts it into a format suitable for further NLP processing. The \textit{automated label generation} component generates text labels from the processed outputs of the \textit{text preparation} method. The method offers two techniques for generating text labels that exploit real, publicly available (human) knowledge such as related scientific works and controlled vocabularies. Finally, the generated text labels from the \textit{automated label generation} step are passed to SciKey's \textit{keyword evaluation} sub-module to assess the quality of the ML-generated keywords. The output from Scikey's \textit{keyword evaluation} are a set of quantitative metric scores that provide information about the quality of the ML-generated keywords when validated against the text labels. It should be noted that the automated label generation component and the first two SciKey modules (\textit{pre-processing} and \textit{keyword extraction}) can be run in parallel. 

\subsection{Text Preparation}\label{methodology-textprep}

\begin{table}
\centering
\caption{List of metadata fields available. The fields containing semantic information are shown in the left column.}
\label{table: metadata_fields}
\begin{tabular}{l|ll}
\hline 
\multicolumn{1}{l}{Semantic fields} & \multicolumn{2}{l}{Non-semantic fields}\tabularnewline
\hline 
\multicolumn{1}{l}{} & Proposal ID & Other Collaborators\tabularnewline
\multicolumn{1}{l}{} & Type & All Collaborators\tabularnewline
\multicolumn{1}{l}{} & Cycle & Survey Comment\tabularnewline
Title & Focus Area & Contracts Office Contact for UA\tabularnewline
Description & PI Name & Transfer Agreements Contact\tabularnewline
Justification & PI Institution & Non-U.S. Samples\tabularnewline
Utilization & PI Email & Samples Regulatory Compliance\tabularnewline
Community Interest & Co-PIs & Primary Funding Source\tabularnewline
DOE Mission & Status & Funding Source Comment\tabularnewline
Sample Preparation & Survey Choice & Completion Date\tabularnewline
Summary of Work & All Institutions & Planned Publications\tabularnewline
\multicolumn{1}{l}{} & Created At & Syn Bio Total KB\tabularnewline
\multicolumn{1}{l}{} & Submitted At
 & Syn Bio Data Mining\tabularnewline
\multicolumn{1}{l}{} & 
Date Approved & \tabularnewline
\hline 
\end{tabular}
\end{table}

The input data for the metadata extraction process is a raw, unlabeled dataset of 2143 approved genomic research proposals available as comma-separated text (\textit{.csv}). The \textit{.csv} file contains a subset of all research proposed for investigation with the facilities available at the JGI over a period of 12 years (2009 - 2020). As shown in Table \ref{table: metadata_fields}, the proposal file contains 35 metadata fields which are predominantly text-based, with only one numerical field (\textit{proposal ID}) and three DateTime fields (containing metadata related to proposal submission and approval dates).

The metadata fields may be classified as containing two types of information:
\begin{enumerate}
\item Fields containing document-related information about the proposal such as author institutions, proposal cycle, and completion date.
\item Fields containing semantic information related to the actual proposed research such as work description, justification, and community interest. 
\end{enumerate}
While the document-related information is useful for provenance and data management purposes, they contain no information of semantic value and only introduce noise to the NLP process. Thus, for the label generation process, we only consider the fields containing semantic information; fields containing only document-related information (excluding the \textit{proposal ID} field) were discarded. In the  \textit{text extraction} step we identify and extract the columns containing relevant semantic information from each proposal. Of the 35 fields available in the raw proposal dataset, eight fields were found to contain useful semantic information about the proposal: the \textit{title, description, justification, community interest, summary of work, sample preparation, utilization} and \textit{DOE mission} fields. 

The text strings contained in the eight fields are joined together to form a single text string for the next steps of the process. 

\subsection{Automated Label Generation}\label{methodology-core}
We implemented two semi-automated techniques for generating quality text labels to be used in validating the ML-generated keywords: artifact linkages and ontologies.

\subsubsection{Artifact linkages} \label{methodology-artifact-linkages}

Linking unlabeled artifacts to directly related artifacts with known labels can provide a set of \emph{derived labels} with which the unlabeled artifact can be associated and/or archived. In our case, we linked the unlabeled proposals to publication records, with each proposal inheriting the keywords from the publication(s) it could be associated with directly. An advantage of linking to publications is that since they contain human labels, the keywords transferred to the unlabeled texts as labels naturally incorporate the necessary semantic knowledge.

\begin{figure}
\begin{minipage}[t]{0.45\columnwidth}%
    \centering
    \includegraphics[scale=0.8]{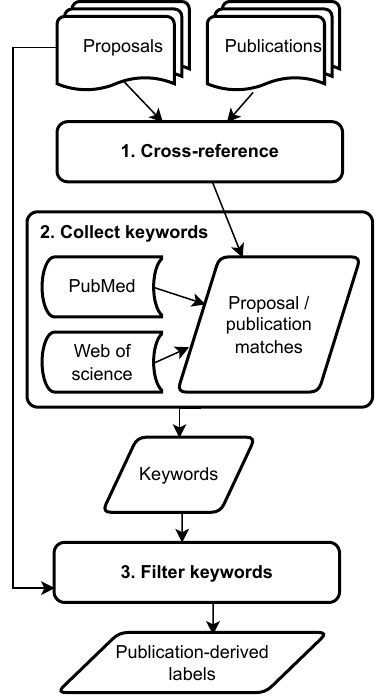}
    \caption{Flowchart of artifact linkage process for label generation}
    \label{fig:art_linkage_flowchart}
\end{minipage}
\hfill{}%
\begin{minipage}[t]{0.45\columnwidth}%
    \centering
    \includegraphics[scale=0.85]{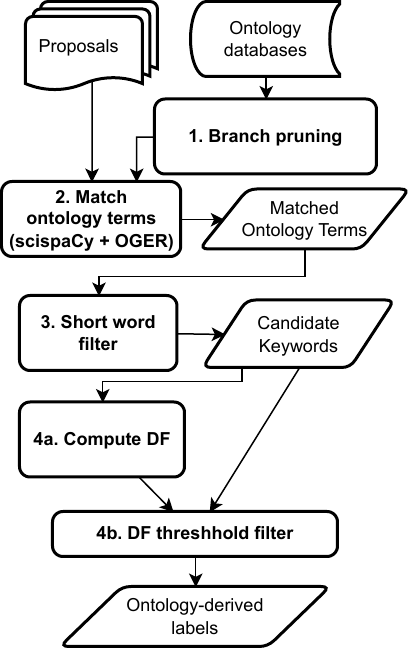}
    \caption{Flowchart of ontology-based process for label generation}
    \label{fig:ont_flowchart}
\end{minipage}
\end{figure}

Figure \ref{fig:art_linkage_flowchart} presents a schematic representation of the artifact linkage process. For our use case, artifact linkage was achieved by cross-referencing the list of proposals against a curated list of publications. In JGI's systems, publications with JGI users and personnel as authors are linked to proposals that produced data or materials used in a given publication. These linkages are established by a combination of automatic assignments and manual curation by JGI staff. The proposal ID field, common to both proposals and publications, provided us with a way to link both types of artifacts.

%Thus, we exploit the fact that the same repository system contains both the proposals and publications related to the Joint Genome Institute. A comparison of the publication metadata available on JAMO with our proposal dataset revealed that some publications were archived with metadata related to the proposal ID, thus allowing us to create a direct link between those publications and specific proposals.

% \begin{figure}
%     \centering
%     \includegraphics[scale=0.8]{artifact-linkages-flowchart-v3.pdf}
%     \caption{Flowchart of artifact linkage process for label generation}
%     \label{fig:art_linkage_flowchart}
% \end{figure}

We created direct links between 184 proposals and 337 publications by cross-referencing the full proposal set against a list of 488 JGI publications. The direct links were created by matching the unique \emph{Proposal ID} field present in both types of data artifacts.
%Cross-referencing the proposal set against a list of 488 JGI publications stored on JAMO system, we were able to create direct links between 184 proposals and 337 publications using the unique \emph{Proposal ID} field present in both data artifacts. 
These 184 proposals with associated publications were considered the training data/subset for the NLP keyphrase extraction model.

%Once the links were created, metadata available for the publications can be easily be acquired and transferred to the proposals. In this case, the publication keywords were retrieved from online scientific databases and search engines. 
For each proposal, the keywords associated with the linked publications were then automatically curated from three online sources:  
\begin{itemize}
\item the \textit{Author keywords} field from Web of Science\footnote{\url{https://clarivate.com/webofsciencegroup/solutions/web-of-science/}}, a collection of keywords chosen by the author of the publication;
\item the \textit{Keywords} field from PubMed\footnote{\url{https://pubmed.ncbi.nlm.nih.gov/}}, also a  collection of author-defined keywords for the publication; and
\item PubMed's \textit{MeSH terms} field, a collection of controlled vocabulary (medical subject headings) used to label articles topically by trained indexers.
\end{itemize}
Users will typically search for documents based on words expected to be present in its contents. Important keywords for a document will typically appear in the text body; words absent from a document's content are unlikely to suitable for searching and indexing it. As such, we filtered out any keywords that did not appear in the proposal text. The remaining keywords obtained from these sources were assigned as training labels for the associated proposals (known as publication-derived labels henceforth).

The artifact linkage approach takes advantage of the relationships between artifacts; it is generally applicable and can be applied to any case where connections between labeled and unlabeled artifacts can be established in some way. This connection can be in the form of numeric tags (e.g., IDs, funding award numbers), strings (e.g., filenames), or even established manually (through interactions with the researchers). Additionally, in research environments, most unlabeled artifacts such as proposals and theses often lead to publications, which can serve as at least one recognized source of labels that is common to all scientific domains. Generally, different artifact types will be related to one another to varying degrees, and the level of relatedness of the artifacts being linked will have an impact on the strength and validity of the keyword association. In our case, there is a clear and direct link between the proposals for a work and the publications that arise out of it, thus the derived labels are mostly expected to have a high degree of validity.

\vspace{5pt}

\subsubsection{Ontology-based text annotation}
\label{methodology-ontologies}
We use expert-curated ontologies to annotate the proposals and identify words/phrases within each unlabeled proposal that are representative of the ideas and topics explored. 

Figure \ref{fig:ont_flowchart} represents the key steps in the ontology-based labeling process. We identify potential metadata using ontologies via a two-stage process:
\begin{enumerate}
\item Identifying relevant phrases and keywords that are relevant to the domain (numbered 1-3 on Fig. \ref{fig:ont_flowchart}).
\item Ranking the identified words and phrases in terms of importance to determine the proposal labels.
\end{enumerate}

% \begin{figure}
%     \centering
%     \includegraphics[scale=0.8]{ontologies-flowchart-v3.pdf}
%     \caption{Flowchart of ontology-based process for label generation}
%     \label{fig:ont_flowchart}
% \end{figure}

\noindent\textbf{Identification of relevant phrases:} 
Written texts such as proposals are typically a mixture of both generic and domain-specific words and phrases. The goal of this step is to identify the list of all candidate labels for each proposal based on information curated by experts with domain knowledge. Ontologies predominantly contain domain-specific terms and phrases, and exploiting them allows us to identify which words present in the proposal domain experts believe are relevant in the context of the environmental genomics domain. 

For our use case, to generate the candidate set of potential text labels for the 184 training proposals, we created an application ontology called BERO (Biological and Environmental Research Ontology), consisting of the genomic, biological, and environmental subject areas and compiled a list of all the identified words and phrases (\textit{matched terms}). Table \ref{table: ontology_list} provides the components ontologies used to create BERO. The ontologies are all open-source and publicly curated. Identified in collaboration with topic experts from JGI, the ontologies cover the entire spectrum of focus areas and work investigated by the institute, including genomics, multiomics, bioinformatics, plants, organisms, biological and environmental entities. We implement the ontology search and entity recognition step by embedding links to the ontologies into text processing and annotation tools specific to the biomedical domain. We use two tools: 
\begin{enumerate*}
\item OntoGene Entity Recognition (OGER) \cite{Basaldella2017, Furrer2019}, a biomedical named entity recognizer, and 
\item scispaCy \cite{Neumann_2019}, a python package for biomedical text processing and Named Entity Recognition (NER). 
\end{enumerate*}
OGER and scispaCy parse the unlabeled text, query the various ontologies and return an annotated list of matched terms. Part-of-speech tagging was done on the text with ScispaCy, allowing us to filter out parts of speech and matched terms that provided no information of semantic value (e.g geographical locations). 

\begin{table}
\centering
\caption{Components of BERO.}
\label{table: ontology_list}
\begin{tabular}{p{6cm} l c} 
\toprule
Ontology & Focus/Domain & Refs. \\ [0.5ex] 
\midrule
 Environment Ontology (\href{https://obofoundry.org/ontology/envo.html}{EnvO}) & Environmental features, habitats & \cite{Buttigieg2013, Buttigieg2016}  \\ 
 Gene Ontology (\href{http://geneontology.org/}{GO}) & Biological functions \& processes & \cite{Ashburner2000, Carbon2020} \\
 Chemical Entities of Biological Interest (\href{https://www.ebi.ac.uk/chebi/}{ChEBI}) & Molecular entities & \cite{Hastings2015} \\
National Centre for Biotechnology Information Taxonomy (\href{https://www.ncbi.nlm.nih.gov/taxonomy}{NCBITaxon}) & Organisms & \cite{Federhen2011} \\
 Ontology of bioscientific data analysis and management (\href{https://edamontology.org/page}{EDAM})  & Bioscientific data \& bioinformatics & \cite{Ison2013} \\
Plant Ontology (\href{https://obofoundry.org/ontology/po.html}{PO})  & Plant anatomy \& genomics & \cite{Cooper2012, Cooper2017} \\
Molecular Process Ontology (MOP) & Molecular processes & \textbf{\cite{Batchelor2020, MOP}} \\
Ontology for Biomedical Investigations (\href{http://obi-ontology.org/}{OBI})  & Biomedical investigations & \cite{Bandrowski2016} \\
Phenotype And Trait Ontology (\href{https://www.ebi.ac.uk/ols/ontologies/pato}{PATO}) & Phynotype qualities & \cite{Gkoutos2017, Gkoutos2004} \\
Ontology of core ecological entities (\href{https://github.com/EcologicalSemantics/ecocore}{ECOCORE}) & Ecological entities & \cite{Ecocore} \\[1ex] 
\botrule
\end{tabular}
\end{table}

An unguided ontology search would return every match found in the proposal texts without taking into account any sort of context, leading to some spurious word and phrase matches. At least two types of spurious matches were found to occur frequently:

\begin{enumerate}
    \item cases where the ontologies matched words or phrases in the proposal exactly, but in the wrong context. This was found to be the case with words that have both domain-specific and general-purpose meaning (e.g., data, well, sample). 
    \item cases in which words in the proposals were wrongly matched to acronyms for domain-specific phrases. This was found to occur with shorter words, especially when word stemming is applied. For example, an unguided search with the word \textit{serv} (the stemmed version of the words serve and service) is a match (and acronym) for  \textit{simian endogeneous retrovirus type D, SERV} in NCBITaxon. 
\end{enumerate}

It is therefore important to implement search and downselection rules to minimize the likelihood of such spurious matches as candidate text labels. To achieve this and keep the size of the matched candidates manageable, two downselection rules were applied:

\begin{description}

\item [Branch pruning:] terms in the "branches" of the ontologies were selectively removed. Concepts in ontologies are typically categorized under a small number of sub-classes called branches; in this case, we only considered a carefully curated selection of branches. This process, called \textit{branch pruning}, was carried out before the ontology search (step 1 in Fig. \ref{fig:ont_flowchart}). The curation process was handled by a domain scientist familiar with the ontology databases to ensure that only relevant ontology sub-classes are retained. For example, within the Ontology of bioscientific data analysis and management (EDAM), of the branches \textit{Topic}, \textit{Operation}, \textit{Data}, \textit{Data Identifier}, and \textit{Format}, only the \textit{Topic} branch  was retained because it includes broader interdisciplinary concepts from 
the biological domain. Similarly, for Phenotype And Trait Ontology (PATO), only the \textit{physical quality} branch was retained. 

\item [Exclude short words:] Words with less than three characters were dropped from the matched terms list. This rule was implemented such that acronyms are unaffected, so important keyphrases like DNA and SOB (Sulphur-Oxidizing Bacteria) are retained. Short word removal is a post-processing step after generating the ontology matches (step 3 in Fig. \ref{fig:ont_flowchart}).

\end{description}

\vspace{3pt}

The result of this step is the set of candidate labels present in the proposal that have been curated by experts with domain knowledge specific to the genomics field. 

\vspace{5pt}

% Once the training proposals have been pre-processed and assigned derived labels from both approaches, keyphrase extraction can proceed. Before passing the sanitized text to the NLP pipeline, word stemming is performed.

\textbf{Ranking and Filtering for Ontology comparisons:} The output of the Ontology search process described above is a rich corpus of words per document. However, the returned words and phrases are unranked and have no associated scores to reflect the relative importance of the different words. Thus, we needed to devise an approach to rank the extracted labels. For this we adopt the \textit{document frequency} (DF), a measure of the rarity of a phrase in a given corpus. The DF for any phrase $w$ is given by:
\[
\textit{DF}(w)=\frac{\text{Number of documents containing the term }w}{\text{Total number of documents in corpus}}
\]

With this metric, the more unique a word is, the more important it is. The frequency of a word is inversely related to its value, with uniqueness treated as a proxy for importance. 

The ontology-based annotation process typically results a large number of word/phrase matches per document. However, documents are typically indexed by a limited set of keywords (typically under 20). Therefore, there needs to be a down-selection of the number of labels based on their importance, as the number of labels can significantly impact the keyword evaluation process. To control the number of text labels to be considered per document, we tested different threshold limits for the DF across the corpus. For any specific threshold limit, only words with a DF score below that limit were considered as text labels for the documents. The threshold limit considered ranged from 1\% (limiting to words occurring at most in two documents in the corpus) to 100\% (no limit on the frequency of occurrence), with 1\% arbitrarily selected as the baseline value. Setting threshold limits achieves two purposes:
\begin{enumerate}
    \item it serves as a way to control the size/number of ontology-derived text labels used in the evaluation process, and
    \item it provides a way to independently assess the performance of the NLP algorithm on different types of ontology-derived text labels. The DF metric is a measure of how unique a keyphrase is, so low DF threshold limits allow us to investigate performance on document-specific labels, while high DF thresholds allow us to evaluate performance on both specific and generic labels.
\end{enumerate}

The matched terms which fall below the set threshold limit were treated as potential labels for the proposals (called ontology-derived labels henceforth). These labels are forwarded to \textit{SciKey's}  keyword evaluation module.

\vspace{5pt}

The only requirement for the ontology-based approach is the availability of the domain-specific vocabularies. It is therefore generalizable to most ontology-aware domains (i.e., domains where collections of controlled vocabularies exist). We believe that there is a sufficiently similar usage and structure to most online ontologies that would allow our methods to apply to new domains. The applicability of the ontology-based approach is expected to  cover a wide breadth of domains, from biology to environmental sciences to linguistics to computing.

\vspace{5pt}
The derived labels generated from the artifact linkage and Ontology-based approaches proposed here are passed to \textit{SciKey} as ground truths to validate the quality of the ML-generated keywords and tune the NLP models (Fig. \ref{fig:pfd-autolabels}).

\subsection{\textit{SciKey} Configuration for NLP keyphrase Extraction and Evaluation}\label{evaluation}

As previously shown in Figure \ref{fig:pfd-autolabels}, the overall goal of the automated label generation process is to provide ground truth labels for the the validation of ML-generated keywords from NLP algorithms. The ML keyword generation and validation process was done using the \textit{SciKey} pipeline. Here, we summarize ML keyword evaluation process with the SciKey pipeline (Figure \ref{fig:scikey_pipeline}) for our use case.

\vspace{5pt}

\noindent\textbf{Pre-processing:} 
First, the text from the \textit{text preparation} step was sanitized for NLP ingestion. Punctuations, URLs, numbers and citations were removed using Regex-based approaches. We employed Named Entity Recognition (NER) techniques and custom expert-curated lists to handle non-standard scientific words and concepts. The sanitized text was then passed to the \textit{keyphrase extraction} module.

\noindent\textbf{Keyword extraction:} For our use case, we selected the YAKE NLP algorithm \cite{Campos2020}; an open-source Python implementation is available on GitHub\footnote{\url{https://github.com/LIAAD/yake}}. While \textit{SciKey} offers other unsupervised learning algorithms available for keyphrase extraction, we focus on YAKE because it showed the best performance among the evaluated algorithms. However, any of the algorithms available in the pipeline could have been selected for the analysis.
YAKE returns the extracted keywords for each proposal. These keywords (called machine-generated or YAKE-generated keywords henceforth) were forwarded to the keyword evaluation component of the ScienceSearch pipeline.  

We used the training subset of  184 proposals to tune the YAKE parameters that control how the ML algorithm was applied, or \emph{hyperparameters}:

\begin{itemize}
\item \textbf{n-gram size}. Longest contiguous sequence of n-words occurring in the text $\left({ngram}\in[1, 2, 3]\right)$, 
\item \textbf{window size}. Sliding window size for YAKE $\left(ws\in[1,2,3]\right)$
\item \textbf{Deduplication method}. Similarity metric for controlling deduplication ($dedup_{m} \in$ [Levenshtein distance, Sequence matcher, Jaro-Winkler]).
\item \textbf{deduplication threshold}. Allowable similarity between candidate ML keyphrases $\left(dedup_v\in[0.6, 0.7, 0.8, 0.9, 0.95]\right)$. 
\end{itemize}

Hyperparameter tuning was done independently for the two sets of derived labels: for each top-$N$ case, with $N$ in $(5,10,20)$, we ran all 135 combinations of the four hyperparameters and chose the combination with the best F-1 scores for the training subset (Eq. \ref{f1_eq}), resulting in a different set hyperparameters in each case. These optimized models could then be used to generate keywords for the 1959 available proposals not in our training subset.

\vspace{5pt}
\noindent\textbf{Keyword Evaluation:}
The quality of the ML-generated keywords generated in the previous stage were evaluated here using the \textit{keyword evaluation} component of the SciKey pipeline. The keyword evaluation module takes two inputs (Fig. \ref{fig:scikey_pipeline}):

\begin{enumerate}
\item a list of machine-generated keywords (produced in the \textit{keyword extraction} sub-module of SciKey), and
\item a list of derived or ground truth labels for the ML labels to be compared against (produced by either the artifact linkage or ontology-based text annotation techniques described in Section \ref{methodology-core}).
\end{enumerate}

We adopt the exact matching approach where the ML-generated keywords are compared against the derived labels for an exact string matching \cite{Papagiannopoulou2019}. For quantitative evaluation, we adopt the classical evaluation metrics used in information retrieval: precision, recall and F-1 \cite{Papagiannopoulou2019},
\begin{align}
    \text{Precision} = \frac{\text{Number of correctly matched keywords}}{\text{Total number of extracted keywords}} &&
\end{align}
\begin{align}
    \text{Recall} = \frac{\text{Number of correctly matched keywords}}{\text{Total number of assigned/derived labels}} &&
\end{align}
\begin{align} 
\text{F-1}=\frac{2}{\text{recall}^{-1}+\text{precision}^{-1}}\label{f1_eq}&&
\end{align}

Here, ``correctly matched" means that the ML-generated keyword is also found in the list of derived labels.  Stemming is applied to both the ML and derived labels using NLTK's \texttt{PorterStemmer}\footnote{\url{https://www.nltk.org/_modules/nltk/stem/porter.html}} to eliminate spurious mismatches.
We generate the metrics for top-$N$ ranked (by YAKE) keywords, with $N$ being 5, 10, or 20.

\subsection{Summary}
We have presented two techniques by which labels may be automatically generated for unlabeled scientific texts. Once the semantically-important section of the unlabeled text is identified and extracted, text labels can be (1) transferred over from generated from directly-related research, or (2) generated using expert-curated Ontologies. The derived labels generated by these techniques can be treated as ground truth labels for evaluating the quality of ML-generated keywords (e.g. from the \textit{SciKey} pipeline). The \textit{automated label generation} pipeline is unique, providing us with alternative ways to validate the quality of the ML-generated keywords without the dependence on direct manual human labeling. For scientific texts, exploiting the domain knowledge already available via ontologies and artifact linkages for the labeling are a first a step towards the automation of keyword extraction.
\section{Results}\label{Results}
This work proposes two techniques -- artifact linkages and ontologies -- for validating ML-generated keywords for unlabeled scientific texts. In this section, we assess the characteristics and quality of both techniques.  First, we present an analysis of the text labels generated by both approaches. We follow this with an analysis of the YAKE ML algorithm with respect to both sets of derived labels (Section \ref{nlp_performance_of_pub_kws}).

These results were generated with Python 3.8.13 on a
Thinkpad X1 Extreme running Windows 10 Pro version 21H2
with 32GB RAM and an Intel i7 processor.

\subsection{Analysis of Derived Labels}

Both approaches derived labels for 184 scientific proposals.

\begin{figure*}
\subfloat[Publication-derived labels]
{\begin{centering}
\includegraphics[scale=0.45]{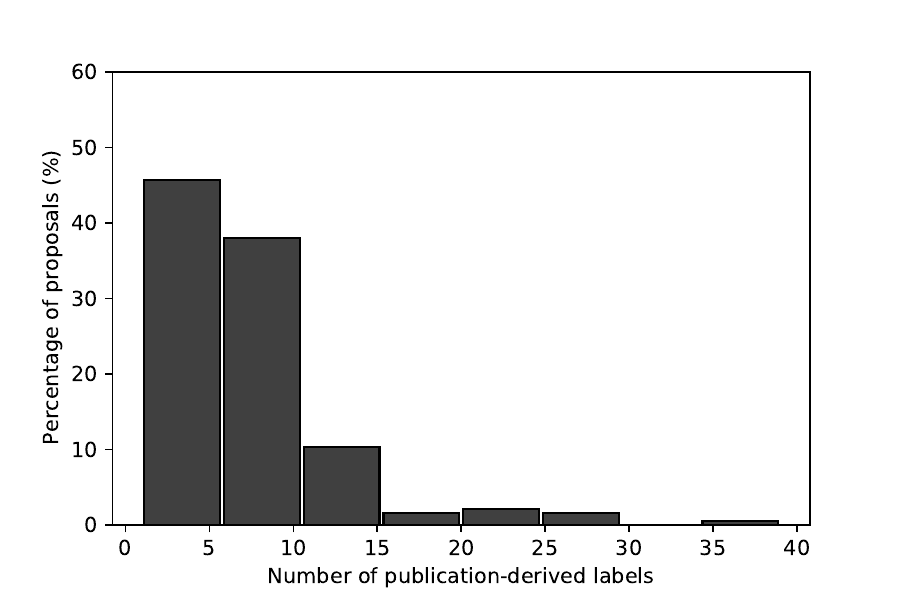}
\label{fig:pub_derived_kws_hist1}
\par\end{centering}
}
\hfill{}
\subfloat[Ontology-derived labels]
{\begin{centering}
\includegraphics[scale=0.45]{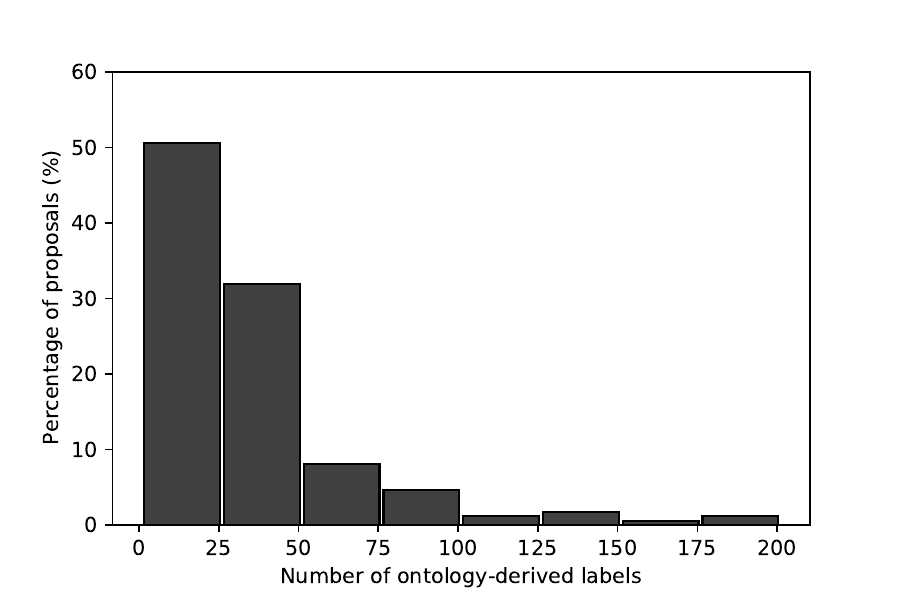}
\label{fig:ont_derived_kws_hist1}
\par\end{centering}
}
\caption{Distribution of derived labels}
\end{figure*}

A total of 1294 labels were obtained from the keywords of publications associated with the proposals, as described in \ref{methodology-artifact-linkages}. Figure \ref{fig:pub_derived_kws_hist1} shows the distribution of the number of labels obtained per proposal. Most of the proposals (83\%) have ten or fewer labels, while roughly 4\% of the proposals have over 20.
%A total of 1294 labels were obtained from the keywords of publications associated with the proposals by creating a direct link to associated publications to proposals -- an average of 7.03 words per proposal.

An assessment of the lengths of the derived labels (Table \ref{tab:kw_lengths_summary}) shows that all the labels had between one word (unigrams) and three words (trigrams). About 85\% of the labels are unigrams (one word), and less than 2\% are trigrams (three words). The result is weighted more towards unigrams which is in line with other works such as ~\citet{Campos2018}, who report averages of 47\%, 34\%, and 13\%  for unigrams, bigrams, and trigrams respectively. 

\subsubsection{Publication-derived labels}\label{analysis_of_pub_kws}

\begin{table}
    \centering
    \caption{\textit{n-gram} summary for derived labels}
    \begin{tabular}{c| r r| r r}
    \hline
     & \multicolumn{2}{c|}{Publication-derived} & \multicolumn{2}{c}{Ontology-derived}\tabularnewline
    \hline
         & No. labels  & Percent (\%) & No. labels  & Percent (\%)\\
         \hline 
    1-gram & 1094 & 84.5 & 3360 & 62.7\\
    2-gram & 183 & 14.1 & 1810 & 33.7\\
    3-gram & 17 & 1.3 & 171 & 3.2\\
    4-gram & & & 22 & 0.4\\
    \hline
    Total & 1294 & & 5363 &\\
    \hline
    \end{tabular}
    \label{tab:kw_lengths_summary}
\end{table}

Table \ref{tab:derived_words_frequency} gives examples of some of the labels found. % across the proposals.
The most frequent labels are typical and representative of the subject area, with the most common keyword, \textit{genom} (stemmed version of genome and genomic), occurring in over 40\% of the proposals.

\begin{table}
    \centering
    \caption{10 most and least common publication-derived labels (after stemming) by frequency}
    \begin{tabular}{p{0.2\linewidth} | p{0.7\linewidth}} 
    \hline
    Most common & 'genom': 76, 'bacteria': 38, 'divers': 36, 'metagenom': 33, 'sequenc': 27, 'gene': 20, 'carbon': 20, 'dna': 19, 'rna': 19,  'soil': 16
           \\
           \hline
    Least common & 'aromat compound': 1, 'valor': 1, 'saccharum': 1, 'hybrid': 1, 'haplotyp': 1, 'phylogenet analysi': 1, 'polyploidi': 1, 'spontaneum': 1, 'sugarcan': 1, 'glycin betain': 1 \\
    \hline
    \end{tabular}
    \label{tab:derived_words_frequency}
\end{table}

\subsubsection{Ontology-derived labels}

A total of 5363 ontology matches were found for the 184 proposals, using the methodology described in \ref{methodology-ontologies}. Figure \ref{fig:ont_derived_kws_hist1} shows the distribution of the number of labels obtained per proposal. Most of the documents ($\approx76\%$) have 50 or fewer labels, while roughly 4\% of the proposals have over 100. An analysis of the ontology terms (Table \ref{tab:kw_lengths_summary}) reveals that the labels are more evenly distributed than in the publication-derived labels case, with bigrams making up just over a third of the labels. Again, very few labels have more than two words.

\vspace{5pt}

The predominance of unigrams and bigrams in both sets of derived labels is in agreement with the conclusion by \citet {Campos2018} that  people rarely use more than three terms to describe a given subject.

\subsection{NLP performance}\label{nlp_performance_of_pub_kws}

We ran the YAKE NLP algorithm and evaluated the generated keywords as described in Section \ref{evaluation}. This section presents the performance against the publication and ontology-derived labels.

\subsubsection{Publication-derived labels}

Table \ref{tab:pub_derkws_exact_match_JGI} shows the results of evaluating YAKE  
%obtained for the JGI proposals with YAKE. 
against the publication-derived labels.
The F-1 scores are similar to recently published results on keyphrase extraction for popular author-labeled scientific datasets such as Krapavin, Semieval2010, NUS, and Inspec (see Table \ref{tab:literature_exact_matching_results}). Thus, we are able to obtain good metadata for the proposals (compared to the state-of-the-art). The best results were obtained at F-1$@ 10$, which indicates that $@10$ provides the best balance between increasing the number of correct matches overall (recall) and keeping the number of false positives low (precision).

\begin{table*}
\caption{YAKE Precision (P), Recall (R), and F-1 scores on derived labels.\label{tab:ert2}}
\subfloat[Publication-derived\label{tab:pub_derkws_exact_match_JGI}]
{
    \begin{tabular}{c r r r}
    \hline
         & P & R & F-1  \\
    \hline
    @5 & 0.255 & 0.246 & 0.250 \\
    @10 & 0.200 & 0.340 & 0.252 \\
    @20 & 0.149 & 0.438 & 0.222  \\
    \hline
    \end{tabular}
}    
\hfill{}
\subfloat[Ontology-derived for DF  threshold of 1\%\label{tab:ont_derkws_exact_match_JGI}]
{
    \begin{tabular}{c r r r}
    \hline
         & P & R & F-1  \\
    \hline
    @5 & 0.094 & 0.105 & 0.099 \\
    @10 & 0.079 & 0.138 & 0.100 \\
    @20 & 0.068 & 0.163 & 0.096  \\
    \hline
    \end{tabular}
}
\end{table*}

\begin{table}
    \centering
    \caption{Best F-1 scores reported in \cite{Papagiannopoulou2019} for some classical scientific collections using the exact matching approach. For comparison, our results are shown in bold on the last row.}
    \begin{tabular}{l | c r r}
    \hline
     Dataset & Type & F-1 @ 10 & F-1 @ 20  \\
    \hline
    NUS & Full-text papers & 0.259 & 0.243 \\
    Krapivin & Full-text papers & 0.190 & 0.161  \\
    Semeval2010 & Full-text papers & 0.208 & 0.219  \\
    Inspec & Paper abstracts & 0.278 & 0.295  \\
    \textbf{Our work (w/YAKE)} & \textbf{Proposals} & \textbf{0.252} & \textbf{0.222} \\
    \hline
    \end{tabular}
    \label{tab:literature_exact_matching_results}
\end{table}

Table \ref{tab:Optimal-YAKE-parameters-pubs} summarizes the optimal YAKE hyperparameter values for the three F-1 cases. The optimal YAKE settings are similar in all cases; the only difference occurs with in the choice of window size for the F-1$@5$ case. The optimal \textit{ngram} size of one is not surprising given the heavy bias of the derived labels towards unigrams, as highlighted in Section \ref{analysis_of_pub_kws}.

\begin{table}
\caption{Optimal YAKE hyperparameters with publication-derived labels}
\label{tab:Optimal-YAKE-parameters-pubs}
\centering{}%
\begin{tabular}{crr>{\centering}m{3cm}r}
\toprule 
 & $ws$ & $ngram$ & $dedup_{m}$ & $dedup_{v}$\tabularnewline
\midrule 
@5 & 3 & 1 & Levenshtein distance, Sequence matcher & 0.9\tabularnewline
@10 & 2 & 1 & Sequence matcher & 0.9\tabularnewline
@20 & 2 & 1 & Sequence matcher & 0.9\tabularnewline
\bottomrule
\end{tabular}
\end{table}

\begin{table}
\caption{Examples of derived and ML-generated keyphrases obtained with YAKE for F-1@10.}

% \begin{tabular}{p{0.02\linewidth} | p{0.35\linewidth} | p{0.25\linewidth} | p{0.3\linewidth}}
\begin{tabular}{>{\raggedright}m{5cm}>{\raggedright}m{3cm}>{\raggedright}m{4cm}}
\hline
Title & Publication-derived labels & ML-generated keyphrases\tabularnewline
\hline
1 -- \textit{Halorespiring Firmicutes: Exploring genomic plasticity of
closely related dedicated degraders with diverging ecophysiological
features and bioremediating capacity} & \textbf{desulfitobacterium}, \textbf{genom} & \textbf{desulfitobacterium}, strain, isol, environment, sequenc, dehalobact,\textbf{
genom}, halorespir, bacteria, degrad\tabularnewline
\hline
2 -- \textit{Resources for study of diversity and divergence in Sorghum,
a C4 cereal model} & polymorph, rice, evolut, diverg, trait, gene,\textbf{ genom},\textbf{
sorghum} & function, saccharum, crop, saccharina, sequenc, grass, variat,\textbf{
genom},\textbf{ sorghum}, divers\tabularnewline
\hline
3 -- \textit{Sequencing the genome of the basidiomycete fungus Amanita
thiersii, a cellulose degrading fungus in an ectomycorrhizal genus} & compar genom, evolut of symbiosi,\textbf{ sequenc},\textbf{ genom},
evolutionari,\textbf{ amanita} & symbiosi, ectomycorrhiz, evolut, thiersii, saprotroph, speci,\textbf{
sequenc},\textbf{ genom}, genu,\textbf{ amanita}\tabularnewline
\hline
\end{tabular}
\label{tab:derkws_exact_examples}
\end{table}

Sample results from the first three proposal documents (Table~\ref{tab:derkws_exact_examples}) illustrate some aspects of the matches that are not obvious from the quantitative metrics. All three of the examples show that we can match both general (e.g. \textit{genom}) and very specific keyphrases (\textit{amanita, desulfitobacterium}). %Comparing the keywords of t
The third document shows that while the ML algorithm is unable to match the trigram keyphrase \textit{evolut of symbiosi} since ``of" is a stopword, 
it does match the component words \textit{evolut} and \textit{symbiosi}.
%we observe the implication of our optimal \textit{n-gram} size of one: the ML algorithm is unable to match the trigram keyphrase \textit{evolut of symbiosi}, %however, the component words \textit{evolut} and \textit{symbiosi}. % are found as independent keywords by the ML algorithm. 
Thus, while some of the keyphrases suggested by the ML algorithm are not present in the exact form listed in the derived labels and are penalized by the exact matching metric, they still represent valid keywords for the documents and should not be discarded. Furthermore, a cursory comparison of the keywords to the proposal titles reveals that some potential representative keywords present in the ML-generated keyphrase list are absent from the derived labels list. For example, the keyword \textit{thersii}  refers to the specific strain of Amanita under investigation and is thus an important keyword; however, the quantitative results do not capture this since it is not present in the list of labels. This suggests that the quality of the metadata obtained for the proposals extends beyond just the keywords matched and quantitative metrics and reflects one of the well known challenges of extracting relevant keywords: the high number of candidate keywords that can be generated from any single text makes it difficult to position the most important ones at the top \cite{Campos2020}.

While the F-1 score is highest for the top-10 keywords, the choice of metrics for a search or indexing task may depend on the specific use case and priorities. In some situations, prioritizing precision may be more appropriate, while in others, maximizing recall is more important. For instance, during research conceptualization, finding all relevant literature on a given subject is crucial, and in such cases, a higher recall may be more desirable. Our results show that predicting the top-20 keywords provides the highest recall, retrieving approximately 44\% of the derived keyphrases (see Table \ref{tab:pub_derkws_exact_match_JGI}). However, if noise in the machine learning (ML) results is a concern, precision may be a more appropriate metric to focus on, and F-1@5 represents the best option, ensuring that 26\% of the generated ML keywords are relevant.

\subsubsection{Ontology-derived labels}\label{nlp_performance_of_ont_kws}

Table \ref{tab:ont_derkws_exact_match_JGI} presents the results obtained for the JGI proposals with YAKE for the baseline DF threshold of 1\%. The F-1 scores obtained are slightly worse than those obtained with the publication-derived labels (Table \ref{tab:pub_derkws_exact_match_JGI}) as well as scientific datasets in literature. This may be attributed to two reasons. First, ontology searches do not account for semantic and contextual information, so while we find many ontology matches, the matches also contain a lot of noise: matches that have domain-specific meaning but are not contextually important. For example, words like \textit{co-culture}, \textit{food}, \textit{annotation}, \textit{assay} and \textit{human} are present as derived labels because they have biological relevance and are thus present in the ontology databases, but they have little value as keywords for the proposals. The absence of ranking for the ontology labels means that these contextually irrelevant matches are difficult to separate out automatically. The NLP algorithm is thus unfairly evaluated because the performance metrics (recall and F-1) are computed based on an inflated number of false negatives. Secondly, ontology databases often have multiple entries for slight variations of the same keyphrase. For example, \textit{dehalobacterium} and \textit{dehalobacterium sp.} exist as different entries in one ontology database and are thus treated as separate labels despite referring to the same bacteria genus.  While a match to either of these terms is sufficient in reality, the evaluation technique penalizes labels not matched exactly (i.e. as false negative), leading to lower F-1 scores. In one specific case, the only derived label returned was \textit{clostridium thermocellum dsm 1313} while the ML algorithm returned \textit{clostridium thermocellum} (as one of its top-5 keywords), leading to a score of zero.

\begin{table}
\caption{YAKE performance on ontology-derived keywords for DF different thresholds.}

\centering{}%
\begin{tabular}{c>{\centering}p{3cm}ccc}
\toprule 
\multirow{2}{*}{Threshold (\%)} & \multirow{2}{3cm}{Average No. Ontology keywords (pre-stemming)} & \multicolumn{3}{c}{F-1}\tabularnewline
\cmidrule{3-5} \cmidrule{4-5} \cmidrule{5-5}
 &  & @5 & @10 & @20\tabularnewline
\midrule
1.0 & 21.64 & 0.099 & \textbf{0.100} & 0.096\tabularnewline
2.0 & 23.50 & 0.102 & \textbf{0.105} & 0.101\tabularnewline
5.0 & 27.20 & 0.107 & \textbf{0.112} & 0.108\tabularnewline
10.0 & 28.48 & 0.108 & \textbf{0.113} & 0.110\tabularnewline
20.0 & 29.39 & 0.108 & \textbf{0.113} & 0.110\tabularnewline
25.0 & 29.55 & 0.108 & \textbf{0.113} & 0.111\tabularnewline
50.0 & 29.88 & 0.109 & \textbf{0.115} & 0.112\tabularnewline
100.0 & 29.93 & 0.109 & \textbf{0.115} & 0.112\tabularnewline
\bottomrule
\end{tabular}
\label{tab:onts_exact_match_JGI}
\end{table}

To understand the sensitivity of the results to the DF threshold value, we re-trained the ML model with different threshold values from 1\% to 100\% and computed the F-1 scores (Table \ref{tab:onts_exact_match_JGI}). The best F-1 scores are obtained when the NLP algorithm generates only ten keywords; further increases in the number of keywords only worsen the NLP algorithm performance. This indicates that keywords that are most unique to each of the proposal texts are ranked high and returned early by the NLP algorithm. The analysis also revealed that precision and recall are influenced differently by the selected threshold and number of keywords.
Precision increases proportionally with the threshold but decreases with higher number of keywords. 
%The magnitude of the impact was found to similar in both cases (28\% precision increase in across threshold range, 34\% precision decrease across the number of keywords). The highest precision obtained was at a threshold value of 100\% for F-1 @ 5. On the other hand, the recall only significantly impacted by the the number of keywords: the recall increases by 116\% when going from F-1@5 to F-1@50. The selected threshold has very little impact on the recall. The highest recall was  obtained at a threshold value of 1\% for F-1 @ 50.
On the other hand, while the recall increases with the number of keywords, it is relatively unaffected by the selected threshold.

In all the F-1@10 cases, the best results were obtained in YAKE with setting \textit{ngram=2}. This marks a change from the optimal setting of \textit{ngram=1} obtained with the publication-derived labels (Section \ref{nlp_performance_of_pub_kws}) and reflects the higher fraction of bigrams in the ontology-derived labels set (Table \ref{tab:kw_lengths_summary}).

\begin{table}
\caption{Examples of ML-generated keyphrases obtained with YAKE @ 10; threshold=1\%. The total number of ontology-derived labels below the threshold is shown in brackets, while the matched keywords are in bold.}

\begin{tabular}{>{\raggedright}m{3.7cm}>{\raggedright}m{5cm}>{\raggedright}m{3.5cm}}
\toprule
Title & Derived labels (from ontologies) & ML-generated keyphrases\tabularnewline
\toprule
1 -- \textit{Halorespiring Firmicutes: Exploring genomic plasticity of
closely related dedicated degraders with diverging ecophysiological
features and bioremediating capacity} &  (36*) - halogen compound,  dehalogen, clostridium, dehalobact restrictu, \textbf{dehalobact}, dehalobact sp.,  verrucomicrobium, gram-posit bacteria, sedimentibact, sedimentibact sp., \textbf{desulfitobacterium}, desulfitobacterium hafniens tcp-a, desulfitobacterium hafniens dp7, desulfitobacterium hafniens, desulfitobacterium metallireducen, threat, reduct, adapt
 &  plastic, elucid genom, halorespir, strain desulfitobacterium, strain, \textbf{dehalobact}, genom sequenc, genom, \textbf{desulfitobacterium}, sequenc \tabularnewline
\hline
2 -- \textit{Resources for study of diversity and divergence in Sorghum,
a C4 cereal model} & (16) - high temperatur, fossil-fuel, population-genet, demograph, strength, mutat rate, bac, gene order, saccharum, water suppli,
 \textbf{johnson grass}, motiv, \textbf{sorghum}, \textbf{saccharina}, bank, attract & saccharina function, sorghum genom, sorghum genu, \textbf{johnson grass}, \textbf{sorghum}, \textbf{saccharina}, cell wall, genom, function genom, sorghum sorghum\tabularnewline
\hline
3 -- \textit{Sequencing the genome of the basidiomycete fungus Amanita
thiersii, a cellulose degrading fungus in an ectomycorrhizal genus} & (7) - decompos, cellulos degrad, amanita , thiersii, laccaria, \textbf{amanita thiersii}, laccaria bicolor, isotop & thiersii genom, genu amanita, compar genom, \textbf{amanita thiersii}, ectomycorrhiz genu, heather hallen, genom sequenc, amanita speci, genom, ectomycorrhiz symbiosi\tabularnewline
\toprule
\end{tabular}
* Only half of the 36 ontology-derived labels are shown here.
\label{tab:ontkws_exact_examples}
\end{table}

Table \ref{tab:ontkws_exact_examples} presents the keywords obtained for the first three proposal documents for F-1@10 and threshold = 1\%. A qualitative comparison of the generated keywords shows that we are able to find unigram and bigram keyphrases specific to each document (e.g. \textit{johnson grass, saccharina}). However, keyphrases that are likely to be more generic (e.g. \textit{genome}) are not validated because they are absent from the ontology-derived label set (eliminated by the frequency filter). For example, the ML-generated keyword \textit{cell wall} is present in the full list of labels for the second proposal, however it occurs five documents (2.7\%) and is therefore not considered in this case. It is worthwhile to note that some of the validated keywords match those found in the publication-derived labels case.

The labels for first document highlights the key challenges with ontology-based approach: we have at least three variants of \textit{desulfitobacterium hafniens}, while some of the labels are would be poor representatives for the proposals (e.g. threat, adapt). Further post-processing of the labels before ML would therefore be beneficial in improving the quality of the results.
\section{Discussion}\label{Discussion}

\noindent{\textbf{Strength of Association and Validity of Derived Labels.}} The ML-generated keywords have been evaluated using keyphrases not directly associated with the texts, thus requiring an assessment of the validity of the derived labels.  The publication-derived labels are user-specified keyphrases from published works that directly leverage data or materials produced through the linked proposals. Thus, these labels are expected to be strongly representative of the proposals, but this may not always be the case. There are two specific conditions where the association may be weak: when the products of the proposal are used but do not play a significant role in the publication, or when the publication covers topics or concepts that differ from those originally stated in the proposal. In such cases, the publication-derived keywords may not accurately reflect the contents of a given proposal. Generally, however, we expect the publication-derived keywords to be representative of the proposals.
The ontology-derived keyphrases are extracted directly from the proposal texts, making them strongly associated. However, with a frequency-based ranking approach for the ontology-derived labels,  questions remain over whether the frequency-based ranks assigned to the individual keyphrases accurately reflect their actual contextual and semantic importance as document labels.  The publication-derived keyphrases are considered the more reliable of the two sources due to the semantic and contextual information incorporated by humans into the keyphrase selection process. Developing a ranking approach that takes into account the contexts of the extracted keyphrases would help increase confidence in the ontology-derived labels.

\noindent{\textbf{Quality and Human-in-the-loop for Derived Labels.}} 
Assigning labels and/or keywords to any document is inherently subjective. User-specified labels are considered the gold standard for text summarization; however, even with that approach, not all the potentially correct keywords are assigned by users. Researchers typically pick keywords in ad-hoc ways that are far from optimal and usually biased \cite{King2017}, and some phrases that are unsuitable as keywords are often included. We encountered the same challenge with the labels generated via artifact linkages (i.e., the publication-derived labels): the ML algorithm found several good candidate keywords that were absent from the derived labels list (e.g., \textit{thiersii}, \textit{halorespir}). Thus, a \textit{post-ML} step of human-in-the-loop keyphrase validation of the ML-generated keywords will be beneficial for improving the quality of the publication-derived labeling approach and ensuring that good keywords are not lost. 

As expected, the ontology search returns significantly more derived keywords per document than the artifact linkage approach. However, it also returns a few generic, low-quality, non-domain specific terms (e.g., \textit{threat}, \textit{strength}, \textit{attract}). With the ontology-derived labels the challenge is the opposite of that described above with artifact linkage, with potential candidate keywords possibly being lost at the filtering stage. Thus, with the ontology-based approach, human-in-the-loop intervention to improve keyword quality will be most beneficial as a \textit{pre-ML} step.

\noindent{\textbf{Generality of ML-generated Keywords.}}
Our results show that the ML-generated keywords for both cases contain some keywords too generic to be semantically or contextually useful (e.g \textit{grass}, \textit{diversity}, \textit{divergence}). These could be eliminated by improving the stopwords list. 

Regarding the validated (i.e matched) keywords, the results show that we can match both generic and document-specific keywords irrespective of the derived labels source. However, the ontology-based approach has the advantage of having a hyperparameter (i.e. the threshold limit) that controls the uniqueness of the validated keywords. This is very useful for eliminating words like \textit{genome} and \textit{sequencing} that, while domain-specific, will occur frequently in a genomic corpus.

\noindent{\textbf{Sensitivity to Stopwords.}} The results observed with the ontology-derived labels were found to be very sensitive to the list of stopwords. Some of the words generally found in scientific texts such as \textit{observations}, \textit{findings} and \textit{field} have domain-specific connotations in the field of genomics and therefore have entries in biological and biomedical ontologies. Such words need to be handled explicitly to avoid their inclusion as potential ontology-derived labels, and the most logical approach is to include them in the list of stopwords. Thus, careful curation of the stopwords list is crucial if the ontology-based labeling process is to be adopted. 
Since the publication-derived labels are human-curated, such words are less likely to exist in the keywords list.

\noindent{\textbf{Performance.}}
The results show that we perform better in validating the labels generated via artifact linkages (i.e., the publication-derived labels generated as described in Section \ref{methodology-artifact-linkages}). This occurs because the publication-derived labels are fewer but of higher quality.  44\% of the ML-generated keywords were validated via the artifact-linkage approach, compared to the 23\% validated with the ontology-derived labels. The two cases require different n-gram settings, each reflecting the \textit{ngram} distribution of their derived labels. Both approaches have some validated keywords in common; such keywords are expected be good representative summarizations of the proposal texts.

\noindent{\textbf{Generality of Proposed Text Labeling Approaches.}}\label{generality-discussion} While this work demonstrates the applicability of the text labeling approaches proposed in the genomics domain, we believe that the techniques are generalizable to other scientific domains.

The artifact linkage approach comprises three main steps (Figure \ref{fig:art_linkage_flowchart}): (1) document cross-referencing, (2) keyword collection for linked artifacts from online scientific databases, and (3) keyword filtering. The cross-referencing step involves the use of additional input in the form of labeled artifacts that can be linked to the unlabeled input texts. Most unlabeled scientific texts can be directly linked to specific scientific research projects that produce other labeled artifacts, such as publications or DOE technical reports. However, while JGI has implemented workflows and practices for associating publications with user proposals, doing so may be a non-trivial and expensive process, and the information required to create these links reliably may not always be available. Further, the criteria for associating research artifacts like publications and proposals could be different at other organizations than those used by JGI. Having access to an existing corpus or a means for producing similar artifact linkages is thus a pre-condition for this approach. Provided that an organization can satisfy this pre-condition, a new application will simply require custom scripts based on how the artifact links are created and the systems involved. The resulting information remains the same as that leveraged in the present study, making the cross-referencing step feasibly generalizable to other domains. Additionally, the cross-referencing step only requires enough information that allows the labeled text to be found (e.g., title, DOI, or PMID), not the full document itself. The second stage of the artifact linkage process is generic and can be applied out of the box if the labeled documents are on PubMed or Web of Science. For other online databases (e.g., Scopus), custom methods specific to parsing information for those databases will be required, but the general concepts will remain the same. The keywords are extracted from the metadata that third-party databases have indexed about the articles, rather than from the articles themselves. The approach, therefore, avoids paywall bottlenecks. The final keyword filtering stage utilizes Regex string matching - it is generic and requires no additional customization to use. Thus, while the process of identifying the link between the labeled and unlabeled artifact may be bespoke based on how connections between the artifacts are established, the other steps are generic.

The ontology-based approach for the text label generation consists of four main steps, as shown in Figure \ref{fig:ont_flowchart}: branch pruning, ontology term matching, short word filter, and threshold filtering. To use the ontology-based approach in a different application area, the first stage of the pipeline needs to be modified to fit the specific domain of interest by incorporating and integrating the links to the appropriate domain-specific ontologies into a Natural Language Processing (NLP) tool. Most scientific application areas are ontology-aware \cite{noy2001}, and there is a reasonable similarity in usage and structure to most online ontologies that would allow the proposed methods to be applied directly to those domains. Thus, while the ontology embedding stage requires some domain-specific customization, the requirements, information and tools required are the same irrespective of domain, making the approach generalizable. The remaining stages of the pipeline require no modifications or domain-specific customizations and can be implemented essentially as described in this work.

%For example, the filtering steps simply down-select the labels purely based on word length and corpus frequency. 

%Embedding an ontology into sciSpaCy is straightforward and requires almost no changes to the existing code beyond editing the ontology links. For non- While we chosen to embed the ontology links into domain-specific scispaCy in this work, exactly the same approach can be used to embed ontology links into the more generic version of the tool, SpaCy \cite{spacy2}. 
\section{Related Work} \label{RelatedWork}

In this section, we present a brief review of works related to keyword extraction, and initiatives to automatically label or augment data for natural language processing.

\vspace{3pt}
\textbf{Keyword Extraction:} There is a large body of work focused on addressing the longstanding problem of extracting relevant keywords from scientific data. NLP provides the capacity to understand \cite{wu2012, mcdonald2016}, summarize \cite{Nasar2019}, paraphrase \cite{Hegde2020}, categorize \cite{Onan2016}, and extract key terms and phrases \cite{Rinartha2021} from scientific texts. Supervised, semi-supervised and unsupervised machine learning methods have all been applied to keyphrase extraction problems to varying degrees of success; \cite{papagiannopoulou2021} and \cite{Hasan2014} provides an extensive review of the current state-of-the-art. 
%Supervised methods have been established as the go-to method for keyword extraction when large collections of manually annotated documents are available \cite{Campos2020}, while unsupervised  techniques have shown great promise for cases where the training corpus is limited. The techniques adopted for keyphrase extraction from scientific documents include graph-based methods \cite{Rinartha2021, Li2017}, statistical methods \cite{ElBeltagy2009}, embedding-based approaches \cite{BennaniSmires2018}, clustering-based methods \cite{Alrehamy2017} and deep-learning \cite{Chen2018, Meng2017, Kim2020}. 
Within the context of keyphrase extraction, the approach we adopt in this work (via SciKey) is unsupervised, with additional capabilities for incorporating domain-specific text processing, named entity recognition, and frequency analysis.

\vspace{3pt}
\textbf{Automated Text Labeling and Data Augmentation:} There have been numerous efforts to automatically label and/or improve the training data for these NLP problems. The traditional approach to addressing the lack of labels has been to focus on generative data augmentation using pre-trained or large language models \cite{Lee2021, Kumar2020, Juuti2020, Papanikolaou2020}. With this approach, a small amount of labeled data is used to train a language model that produces labeled synthetic data for supervised NLP tasks, with the synthetic data used to train the final NLP model. For example, several researchers have adopted this approach to perform data augmentation for text classification by fine-tuning a language model to synthesize new inputs $x$ for a given label $y$ \cite{Lee2021, Kumar2020}. Similarly, data augmentation techniques such as \textit{AugGPT} \cite{Dai2023}, \textit{GPT3Mix} \cite{Yoo2021},  \textit{LAMBADA} \cite{AnabyTavor2019} and \textit{DARE} \cite{Papanikolaou2020} generate synthetic training data for supervised learning and text classification by fine-tuning neural network models such as GPT-2 \cite{radford2019} and GPT-3 \cite{Brown2020}. More recently, the concept of \textit{zero-label language learning} for NLPs has been explored to eliminate the need for fine-tuning the pre-trained language models \cite{Wang2021, Brown2020}. With zero-label language learning, no human-annotated data is used anywhere during training: the NLP models are trained purely on synthetic data generated from pre-trained language models. For example, the \textit{Unsupervised Data Generation (UDG)} technique \cite{Wang2021} uses a few user-supplied unlabeled examples to train a language model to synthesize high-quality training data without real human annotations; the model produces comparable results to baseline models trained on human-labeled data for classification problems. 
Furthermore, with the successes of foundational language models such as GPT-3, GPT-4 \cite{OpenAI2023}, and LLaMA \cite{Touvron2023, Touvron2023a}, there have been attempts to apply large language models directly to unlabeled data annotation task via prompt engineering and in-context learning; for example, \cite{Ding2022} evaluates the capabilities of GPT-3 for different data annotation tasks such as classification and Named Entity Recognition (NER).
%The models trained with the synthetic data were found to produce comparable results to baseline models trained on human-labeled data for classification problems. 

Weakly supervised learning approaches that depend on programmable rules and heuristics (i.e. labeling functions) to generate labels or synthesize new examples have also been explored \cite{Perez2017}, however, the technique works best on classification problems, and defining sensible rules is difficult. 

The above approaches allow us to generate labeled training data for supervised and unsupervised matching learning training and validation; however, they either require some labeled examples to fine-tune the language models \cite{Ding2022}, work only for text classification, do not label the existing data directly, require significant post-processing \cite{Wang2021}, or do not take into account the domain-specific nature of scientific texts, thus limiting their applicability to the labeling of domain-specific scientific artifacts for keyphrase extraction. Using pre-trained language models for text labeling is also computationally
intensive and often fails to cover the full diversity and
complexity of real examples \cite{Lee2021}. Previous research has shown that large language models like GPT-3 may not perform well when applied directly to complex data annotation tasks such as NER without additional fine-tuning \citep{Ding2022, jin2023genegpt}. 

The methods proposed in this work address some of these challenges. Our approaches generate labels for the existing training data rather than creating synthetic training data; they require no pre-trained language models or labeled examples. A unique feature of the methods we present in this work is that our proposed techniques do not use machine learning techniques for the label generation/assignment task at all; instead, we primarily exploit real available public (human) knowledge such as related scientific works and controlled vocabularies. This has several advantages. First, our proposed approaches are particularly suited to handle and exploit the domain-specific nature of scientific texts, a task that would be more difficult with pre-trained general language models. Secondly, in addition to being computationally inexpensive, the methods developed here allow us to carry out keyword assignment and extraction tasks, not just text classification. 
\section{Conclusion}\label{Conclusion}
In this work, we presented two approaches to automatically generating labels for validating ML-generated keywords for unlabeled texts. The first approach overcomes the lack of user-defined labels by exploiting direct links between scientific proposals and publications. In the second approach, we take advantage of domain-specific ontologies and frequency-based techniques to produce a set of derived labels against which the ML-generated keywords were validated. The results show varying degrees of success for both approaches based on the exact matching technique, with up to 44\% of the link-derived keywords found by the ML algorithm, and more than to one in four of the extracted ML keyphrases found to be relevant. The approaches presented in this work can be applied to enhance indexing and improve the search of unlabeled texts in scientific databases and other information retrieval systems.

In the future, we plan to develop additional intelligent techniques for ranking the ontology-derived labels that incorporate semantic and contextual information. Furthermore, we plan to expand these approaches for validating unlabeled texts to other scientific artifacts (reports and theses)  and other domains such as earth sciences.

\backmatter

\section*{Declarations}

\bmhead{Ethics approval and consent to participate}
Not applicable.

\bmhead{Consent for publication}
Not applicable.

\bmhead{Competing interests}
The authors declare that they have no competing interests.

\bmhead{Funding}
This work is supported by the Office of Advanced Scientific Computing Research (ASCR) program and the Joint Genome Institute (\url{https://ror.org/04xm1d337}) and supported by the Office of Science of the U.S. Department of Energy operated under Contract No. DE-AC02-05CH11231.

\bmhead{Authors' contributions}
The summary of the contributions of the individual authors are as follows:

OA: Conceptualization, Methodology development - artifact linkages, Methodology development - ontologies, Software, Data acquisition, Data analysis, Writing - Original Draft, Writing - Review \& Editing, Visualization.

HH: Methodology development - ontologies, Software - BERO ontology, Data acquisition.

CM: Methodology development - ontologies, Software - BERO ontology, Data acquisition, Writing - Review \& Editing.

AG: Conceptualization, Methodology development - artifact linkages, Formal analysis.

NB: Methodology development - artifact linkages, Data acquisition, Formal analysis, Writing - Review \& Editing.

DG: Conceptualization, Writing - Original Draft, Writing - Review \& Editing, Visualization.

KF: Conceptualization, Methodology development, Data acquisition, Writing - Review \& Editing.

LR:  Conceptualization, Methodology development - artifact linkages, Methodology development - ontologies, Writing - Original Draft, Writing - Review \& Editing, Supervision.

All authors read and approved the final manuscript.

\bmhead{Acknowledgements}
This material is based on work supported by the U.S. Department of Energy, Office of Science, Office of Advanced Scientific Computing Research (ASCR) and the Joint Genome Institute, a DOE Office of Science User Facility, both under Contract No. DE-AC02-05CH11231. We would like to thank the ScienceSearch team for their invaluable contributions, useful discussions and technical advice. 

\bmhead{Availability of data and materials}
The dataset supporting the conclusions of this article contains sensitive and confidential research information that, by its nature, cannot be made openly accessible.

\bibliography{main}% common bib file


\begin{thebibliography}{55}
\ifx \bisbn   \undefined \def \bisbn  #1{ISBN #1}\fi
\ifx \binits  \undefined \def \binits#1{#1}\fi
\ifx \bauthor  \undefined \def \bauthor#1{#1}\fi
\ifx \batitle  \undefined \def \batitle#1{#1}\fi
\ifx \bjtitle  \undefined \def \bjtitle#1{#1}\fi
\ifx \bvolume  \undefined \def \bvolume#1{\textbf{#1}}\fi
\ifx \byear  \undefined \def \byear#1{#1}\fi
\ifx \bissue  \undefined \def \bissue#1{#1}\fi
\ifx \bfpage  \undefined \def \bfpage#1{#1}\fi
\ifx \blpage  \undefined \def \blpage #1{#1}\fi
\ifx \burl  \undefined \def \burl#1{\textsf{#1}}\fi
\ifx \doiurl  \undefined \def \doiurl#1{\url{https://doi.org/#1}}\fi
\ifx \betal  \undefined \def \betal{\textit{et al.}}\fi
\ifx \binstitute  \undefined \def \binstitute#1{#1}\fi
\ifx \binstitutionaled  \undefined \def \binstitutionaled#1{#1}\fi
\ifx \bctitle  \undefined \def \bctitle#1{#1}\fi
\ifx \beditor  \undefined \def \beditor#1{#1}\fi
\ifx \bpublisher  \undefined \def \bpublisher#1{#1}\fi
\ifx \bbtitle  \undefined \def \bbtitle#1{#1}\fi
\ifx \bedition  \undefined \def \bedition#1{#1}\fi
\ifx \bseriesno  \undefined \def \bseriesno#1{#1}\fi
\ifx \blocation  \undefined \def \blocation#1{#1}\fi
\ifx \bsertitle  \undefined \def \bsertitle#1{#1}\fi
\ifx \bsnm \undefined \def \bsnm#1{#1}\fi
\ifx \bsuffix \undefined \def \bsuffix#1{#1}\fi
\ifx \bparticle \undefined \def \bparticle#1{#1}\fi
\ifx \barticle \undefined \def \barticle#1{#1}\fi
\bibcommenthead
\ifx \bconfdate \undefined \def \bconfdate #1{#1}\fi
\ifx \botherref \undefined \def \botherref #1{#1}\fi
\ifx \url \undefined \def \url#1{\textsf{#1}}\fi
\ifx \bchapter \undefined \def \bchapter#1{#1}\fi
\ifx \bbook \undefined \def \bbook#1{#1}\fi
\ifx \bcomment \undefined \def \bcomment#1{#1}\fi
\ifx \oauthor \undefined \def \oauthor#1{#1}\fi
\ifx \citeauthoryear \undefined \def \citeauthoryear#1{#1}\fi
\ifx \endbibitem  \undefined \def \endbibitem {}\fi
\ifx \bconflocation  \undefined \def \bconflocation#1{#1}\fi
\ifx \arxivurl  \undefined \def \arxivurl#1{\textsf{#1}}\fi
\csname PreBibitemsHook\endcsname

%%% 1
\bibitem[\protect\citeauthoryear{Weber et~al.}{2018}]{Gunther2018}
\begin{bchapter}
\bauthor{\bsnm{Weber}, \binits{G.H.}},
\bauthor{\bsnm{Ophus}, \binits{C.}},
\bauthor{\bsnm{Ramakrishnan}, \binits{L.}}:
\bctitle{Automated labeling of electron microscopy images using deep learning}.
In: \bbtitle{2018 IEEE/ACM Machine Learning in HPC Environments (MLHPC)},
pp. \bfpage{26}--\blpage{36}
(\byear{2018}).
\doiurl{10.1109/MLHPC.2018.8638633}
\end{bchapter}
\endbibitem

%%% 2
\bibitem[\protect\citeauthoryear{Papagiannopoulou and
  Tsoumakas}{2019}]{Papagiannopoulou2019}
\begin{botherref}
\oauthor{\bsnm{Papagiannopoulou}, \binits{E.}},
\oauthor{\bsnm{Tsoumakas}, \binits{G.}}:
A review of keyphrase extraction.
{WIREs} Data Mining and Knowledge Discovery
\textbf{10}(2)
(2019)
\doiurl{10.1002/widm.1339}
\end{botherref}
\endbibitem

%%% 3
\bibitem[\protect\citeauthoryear{Ding et~al.}{2022}]{Ding2022}
\begin{botherref}
\oauthor{\bsnm{Ding}, \binits{B.}},
\oauthor{\bsnm{Qin}, \binits{C.}},
\oauthor{\bsnm{Liu}, \binits{L.}},
\oauthor{\bsnm{Bing}, \binits{L.}},
\oauthor{\bsnm{Joty}, \binits{S.}},
\oauthor{\bsnm{Li}, \binits{B.}}:
Is GPT-3 a Good Data Annotator?
arXiv
(2022).
\doiurl{10.48550/ARXIV.2212.10450}
\end{botherref}
\endbibitem

%%% 4
\bibitem[\protect\citeauthoryear{Bada et~al.}{2012}]{Bada2012}
\begin{botherref}
\oauthor{\bsnm{Bada}, \binits{M.}},
\oauthor{\bsnm{Eckert}, \binits{M.}},
\oauthor{\bsnm{Evans}, \binits{D.}},
\oauthor{\bsnm{Garcia}, \binits{K.}},
\oauthor{\bsnm{Shipley}, \binits{K.}},
\oauthor{\bsnm{Sitnikov}, \binits{D.}},
\oauthor{\bsnm{Baumgartner}, \binits{W.A.}},
\oauthor{\bsnm{Cohen}, \binits{K.B.}},
\oauthor{\bsnm{Verspoor}, \binits{K.}},
\oauthor{\bsnm{Blake}, \binits{J.A.}},
\oauthor{\bsnm{Hunter}, \binits{L.E.}}:
Concept annotation in the {CRAFT} corpus.
{BMC} Bioinformatics
\textbf{13}(1)
(2012)
\doiurl{10.1186/1471-2105-13-161}
\end{botherref}
\endbibitem

%%% 5
\bibitem[\protect\citeauthoryear{Kumar et~al.}{2020}]{Kumar2020}
\begin{botherref}
\oauthor{\bsnm{Kumar}, \binits{V.}},
\oauthor{\bsnm{Choudhary}, \binits{A.}},
\oauthor{\bsnm{Cho}, \binits{E.}}:
Data Augmentation using Pre-trained Transformer Models.
arXiv
(2020).
\doiurl{10.48550/ARXIV.2003.02245}
\end{botherref}
\endbibitem

%%% 6
\bibitem[\protect\citeauthoryear{Papanikolaou and
  Pierleoni}{2020}]{Papanikolaou2020}
\begin{botherref}
\oauthor{\bsnm{Papanikolaou}, \binits{Y.}},
\oauthor{\bsnm{Pierleoni}, \binits{A.}}:
DARE: Data Augmented Relation Extraction with GPT-2.
arXiv
(2020).
\doiurl{10.48550/ARXIV.2004.13845}
\end{botherref}
\endbibitem

%%% 7
\bibitem[\protect\citeauthoryear{Lee et~al.}{2021}]{Lee2021}
\begin{botherref}
\oauthor{\bsnm{Lee}, \binits{K.}},
\oauthor{\bsnm{Guu}, \binits{K.}},
\oauthor{\bsnm{He}, \binits{L.}},
\oauthor{\bsnm{Dozat}, \binits{T.}},
\oauthor{\bsnm{Chung}, \binits{H.W.}}:
Neural Data Augmentation via Example Extrapolation.
arXiv
(2021).
\doiurl{10.48550/ARXIV.2102.01335}
\end{botherref}
\endbibitem

%%% 8
\bibitem[\protect\citeauthoryear{Brank et~al.}{2005}]{brank2005}
\begin{bchapter}
\bauthor{\bsnm{Brank}, \binits{J.}},
\bauthor{\bsnm{Grobelnik}, \binits{M.}},
\bauthor{\bsnm{Mladenic}, \binits{D.}}:
\bctitle{A survey of ontology evaluation techniques}.
In: \bbtitle{Proceedings of the Conference on Data Mining and Data Warehouses
  (SiKDD 2005)},
pp. \bfpage{166}--\blpage{170}
(\byear{2005}).
\bcomment{Citeseer Ljubljana Slovenia}
\end{bchapter}
\endbibitem

%%% 9
\bibitem[\protect\citeauthoryear{Jackson et~al.}{2021}]{Jackson2021}
\begin{botherref}
\oauthor{\bsnm{Jackson}, \binits{R.}},
\oauthor{\bsnm{Matentzoglu}, \binits{N.}},
\oauthor{\bsnm{Overton}, \binits{J.A.}},
\oauthor{\bsnm{Vita}, \binits{R.}},
\oauthor{\bsnm{Balhoff}, \binits{J.P.}},
\oauthor{\bsnm{Buttigieg}, \binits{P.L.}},
\oauthor{\bsnm{Carbon}, \binits{S.}},
\oauthor{\bsnm{Courtot}, \binits{M.}},
\oauthor{\bsnm{Diehl}, \binits{A.D.}},
\oauthor{\bsnm{Dooley}, \binits{D.M.}},
\oauthor{\bsnm{Duncan}, \binits{W.D.}},
\oauthor{\bsnm{Harris}, \binits{N.L.}},
\oauthor{\bsnm{Haendel}, \binits{M.A.}},
\oauthor{\bsnm{Lewis}, \binits{S.E.}},
\oauthor{\bsnm{Natale}, \binits{D.A.}},
\oauthor{\bsnm{Osumi-Sutherland}, \binits{D.}},
\oauthor{\bsnm{Ruttenberg}, \binits{A.}},
\oauthor{\bsnm{Schriml}, \binits{L.M.}},
\oauthor{\bsnm{Smith}, \binits{B.}},
\oauthor{\bsnm{Jr.}, \binits{C.J.S.}},
\oauthor{\bsnm{Vasilevsky}, \binits{N.A.}},
\oauthor{\bsnm{Walls}, \binits{R.L.}},
\oauthor{\bsnm{Zheng}, \binits{J.}},
\oauthor{\bsnm{Mungall}, \binits{C.J.}},
\oauthor{\bsnm{Peters}, \binits{B.}}:
{OBO} foundry in 2021: operationalizing open data principles to evaluate
  ontologies.
Database
\textbf{2021}
(2021)
\doiurl{10.1093/database/baab069}
\end{botherref}
\endbibitem

%%% 10
\bibitem[\protect\citeauthoryear{Rodrigo et~al.}{2018}]{Rodrigo2018}
\begin{bchapter}
\bauthor{\bsnm{Rodrigo}, \binits{G.P.}},
\bauthor{\bsnm{Henderson}, \binits{M.}},
\bauthor{\bsnm{Weber}, \binits{G.H.}},
\bauthor{\bsnm{Ophus}, \binits{C.}},
\bauthor{\bsnm{Antypas}, \binits{K.}},
\bauthor{\bsnm{Ramakrishnan}, \binits{L.}}:
\bctitle{{ScienceSearch}: Enabling search through automatic metadata
  generation}.
In: \bbtitle{2018 {IEEE} 14th International Conference on e-Science
  (e-Science)}.
\bpublisher{{IEEE}}, \blocation{???}
(\byear{2018}).
\doiurl{10.1109/escience.2018.00025}
\end{bchapter}
\endbibitem

%%% 11
\bibitem[\protect\citeauthoryear{Campos et~al.}{2020}]{Campos2020}
\begin{barticle}
\bauthor{\bsnm{Campos}, \binits{R.}},
\bauthor{\bsnm{Mangaravite}, \binits{V.}},
\bauthor{\bsnm{Pasquali}, \binits{A.}},
\bauthor{\bsnm{Jorge}, \binits{A.}},
\bauthor{\bsnm{Nunes}, \binits{C.}},
\bauthor{\bsnm{Jatowt}, \binits{A.}}:
\batitle{{YAKE}! keyword extraction from single documents using multiple local
  features}.
\bjtitle{Information Sciences}
\bvolume{509},
\bfpage{257}--\blpage{289}
(\byear{2020})
\doiurl{10.1016/j.ins.2019.09.013}
\end{barticle}
\endbibitem

%%% 12
\bibitem[\protect\citeauthoryear{Giannakou et~al.}{}]{Giannakou2024}
\begin{botherref}
\oauthor{\bsnm{Giannakou}, \binits{A.}},
\oauthor{\bsnm{Amusat}, \binits{O.}},
\oauthor{\bsnm{Sanyal}, \binits{D.}},
\oauthor{\bsnm{Ramakrishnan}, \binits{L.}}:
Sci-Key: A Keyword Extraction Pipeline for Scientific Documents.
In preparation, 2023
\end{botherref}
\endbibitem

%%% 13
\bibitem[\protect\citeauthoryear{Mihalcea and Tarau}{2004}]{mihalcea2004}
\begin{bchapter}
\bauthor{\bsnm{Mihalcea}, \binits{R.}},
\bauthor{\bsnm{Tarau}, \binits{P.}}:
\bctitle{Textrank: Bringing order into text}.
In: \bbtitle{Proceedings of the 2004 Conference on Empirical Methods in Natural
  Language Processing},
pp. \bfpage{404}--\blpage{411}
(\byear{2004})
\end{bchapter}
\endbibitem

%%% 14
\bibitem[\protect\citeauthoryear{Rose et~al.}{2010}]{Rose2010}
\begin{bchapter}
\bauthor{\bsnm{Rose}, \binits{S.}},
\bauthor{\bsnm{Engel}, \binits{D.}},
\bauthor{\bsnm{Cramer}, \binits{N.}},
\bauthor{\bsnm{Cowley}, \binits{W.}}:
\bctitle{Automatic keyword extraction from individual documents}.
In: \bbtitle{Text Mining},
pp. \bfpage{1}--\blpage{20}.
\bpublisher{John Wiley {\&} Sons, Ltd}, \blocation{???}
(\byear{2010}).
\doiurl{10.1002/9780470689646.ch1}
\end{bchapter}
\endbibitem

%%% 15
\bibitem[\protect\citeauthoryear{Basaldella et~al.}{2017}]{Basaldella2017}
\begin{botherref}
\oauthor{\bsnm{Basaldella}, \binits{M.}},
\oauthor{\bsnm{Furrer}, \binits{L.}},
\oauthor{\bsnm{Tasso}, \binits{C.}},
\oauthor{\bsnm{Rinaldi}, \binits{F.}}:
Entity recognition in the biomedical domain using a hybrid approach.
Journal of Biomedical Semantics
\textbf{8}(1)
(2017)
\doiurl{10.1186/s13326-017-0157-6}
\end{botherref}
\endbibitem

%%% 16
\bibitem[\protect\citeauthoryear{Furrer et~al.}{2019}]{Furrer2019}
\begin{botherref}
\oauthor{\bsnm{Furrer}, \binits{L.}},
\oauthor{\bsnm{Jancso}, \binits{A.}},
\oauthor{\bsnm{Colic}, \binits{N.}},
\oauthor{\bsnm{Rinaldi}, \binits{F.}}:
{OGER++}: hybrid multi-type entity recognition.
Journal of Cheminformatics
\textbf{11}(1)
(2019)
\doiurl{10.1186/s13321-018-0326-3}
\end{botherref}
\endbibitem

%%% 17
\bibitem[\protect\citeauthoryear{Neumann et~al.}{2019}]{Neumann_2019}
\begin{bchapter}
\bauthor{\bsnm{Neumann}, \binits{M.}},
\bauthor{\bsnm{King}, \binits{D.}},
\bauthor{\bsnm{Beltagy}, \binits{I.}},
\bauthor{\bsnm{Ammar}, \binits{W.}}:
\bctitle{{ScispaCy}: Fast and robust models for biomedical natural language
  processing}.
In: \bbtitle{Proceedings of the 18th {BioNLP} Workshop and Shared Task}.
\bpublisher{Association for Computational Linguistics}, \blocation{???}
(\byear{2019}).
\doiurl{10.18653/v1/w19-5034} .
\burl{https://doi.org/10.18653%2Fv1%2Fw19-5034}
\end{bchapter}
\endbibitem

%%% 18
\bibitem[\protect\citeauthoryear{Buttigieg et~al.}{2013}]{Buttigieg2013}
\begin{barticle}
\bauthor{\bsnm{Buttigieg}, \binits{P.}},
\bauthor{\bsnm{Morrison}, \binits{N.}},
\bauthor{\bsnm{Smith}, \binits{B.}},
\bauthor{\bsnm{Mungall}, \binits{C.J.}},
\bauthor{\bsnm{and}, \binits{S.E.L.}}:
\batitle{The environment ontology: contextualising biological and biomedical
  entities}.
\bjtitle{Journal of Biomedical Semantics}
\bvolume{4}(\bissue{1}),
\bfpage{43}
(\byear{2013})
\doiurl{10.1186/2041-1480-4-43}
\end{barticle}
\endbibitem

%%% 19
\bibitem[\protect\citeauthoryear{Buttigieg et~al.}{2016}]{Buttigieg2016}
\begin{botherref}
\oauthor{\bsnm{Buttigieg}, \binits{P.L.}},
\oauthor{\bsnm{Pafilis}, \binits{E.}},
\oauthor{\bsnm{Lewis}, \binits{S.E.}},
\oauthor{\bsnm{Schildhauer}, \binits{M.P.}},
\oauthor{\bsnm{Walls}, \binits{R.L.}},
\oauthor{\bsnm{Mungall}, \binits{C.J.}}:
The environment ontology in 2016: bridging domains with increased scope,
  semantic density, and interoperation.
Journal of Biomedical Semantics
\textbf{7}(1)
(2016)
\doiurl{10.1186/s13326-016-0097-6}
\end{botherref}
\endbibitem

%%% 20
\bibitem[\protect\citeauthoryear{Ashburner et~al.}{2000}]{Ashburner2000}
\begin{barticle}
\bauthor{\bsnm{Ashburner}, \binits{M.}},
\bauthor{\bsnm{Ball}, \binits{C.A.}},
\bauthor{\bsnm{Blake}, \binits{J.A.}},
\bauthor{\bsnm{Botstein}, \binits{D.}},
\bauthor{\bsnm{Butler}, \binits{H.}},
\bauthor{\bsnm{Cherry}, \binits{J.M.}},
\bauthor{\bsnm{Davis}, \binits{A.P.}},
\bauthor{\bsnm{Dolinski}, \binits{K.}},
\bauthor{\bsnm{Dwight}, \binits{S.S.}},
\bauthor{\bsnm{Eppig}, \binits{J.T.}},
\bauthor{\bsnm{Harris}, \binits{M.A.}},
\bauthor{\bsnm{Hill}, \binits{D.P.}},
\bauthor{\bsnm{Issel-Tarver}, \binits{L.}},
\bauthor{\bsnm{Kasarskis}, \binits{A.}},
\bauthor{\bsnm{Lewis}, \binits{S.}},
\bauthor{\bsnm{Matese}, \binits{J.C.}},
\bauthor{\bsnm{Richardson}, \binits{J.E.}},
\bauthor{\bsnm{Ringwald}, \binits{M.}},
\bauthor{\bsnm{Rubin}, \binits{G.M.}},
\bauthor{\bsnm{Sherlock}, \binits{G.}}:
\batitle{Gene ontology: tool for the unification of biology}.
\bjtitle{Nature Genetics}
\bvolume{25}(\bissue{1}),
\bfpage{25}--\blpage{29}
(\byear{2000})
\doiurl{10.1038/75556}
\end{barticle}
\endbibitem

%%% 21
\bibitem[\protect\citeauthoryear{{The Gene Ontology
  Consortium}}{2020}]{Carbon2020}
\begin{barticle}
\bauthor{\bsnm{{The Gene Ontology Consortium}}}:
\batitle{The {G}ene {O}ntology resource: enriching a {GOld} mine}.
\bjtitle{Nucleic Acids Research}
\bvolume{49}(\bissue{D1}),
\bfpage{325}--\blpage{334}
(\byear{2020})
\doiurl{10.1093/nar/gkaa1113}
\end{barticle}
\endbibitem

%%% 22
\bibitem[\protect\citeauthoryear{Hastings et~al.}{2015}]{Hastings2015}
\begin{barticle}
\bauthor{\bsnm{Hastings}, \binits{J.}},
\bauthor{\bsnm{Owen}, \binits{G.}},
\bauthor{\bsnm{Dekker}, \binits{A.}},
\bauthor{\bsnm{Ennis}, \binits{M.}},
\bauthor{\bsnm{Kale}, \binits{N.}},
\bauthor{\bsnm{Muthukrishnan}, \binits{V.}},
\bauthor{\bsnm{Turner}, \binits{S.}},
\bauthor{\bsnm{Swainston}, \binits{N.}},
\bauthor{\bsnm{Mendes}, \binits{P.}},
\bauthor{\bsnm{Steinbeck}, \binits{C.}}:
\batitle{{ChEBI} in 2016: Improved services and an expanding collection of
  metabolites}.
\bjtitle{Nucleic Acids Research}
\bvolume{44}(\bissue{D1}),
\bfpage{1214}--\blpage{1219}
(\byear{2015})
\doiurl{10.1093/nar/gkv1031}
\end{barticle}
\endbibitem

%%% 23
\bibitem[\protect\citeauthoryear{Federhen}{2011}]{Federhen2011}
\begin{barticle}
\bauthor{\bsnm{Federhen}, \binits{S.}}:
\batitle{The {NCBI} taxonomy database}.
\bjtitle{Nucleic Acids Research}
\bvolume{40}(\bissue{D1}),
\bfpage{136}--\blpage{143}
(\byear{2011})
\doiurl{10.1093/nar/gkr1178}
\end{barticle}
\endbibitem

%%% 24
\bibitem[\protect\citeauthoryear{Ison et~al.}{2013}]{Ison2013}
\begin{barticle}
\bauthor{\bsnm{Ison}, \binits{J.}},
\bauthor{\bsnm{Kalas}, \binits{M.}},
\bauthor{\bsnm{Jonassen}, \binits{I.}},
\bauthor{\bsnm{Bolser}, \binits{D.}},
\bauthor{\bsnm{Uludag}, \binits{M.}},
\bauthor{\bsnm{McWilliam}, \binits{H.}},
\bauthor{\bsnm{Malone}, \binits{J.}},
\bauthor{\bsnm{Lopez}, \binits{R.}},
\bauthor{\bsnm{Pettifer}, \binits{S.}},
\bauthor{\bsnm{Rice}, \binits{P.}}:
\batitle{{EDAM}: an ontology of bioinformatics operations, types of data and
  identifiers, topics and formats}.
\bjtitle{Bioinformatics}
\bvolume{29}(\bissue{10}),
\bfpage{1325}--\blpage{1332}
(\byear{2013})
\doiurl{10.1093/bioinformatics/btt113}
\end{barticle}
\endbibitem

%%% 25
\bibitem[\protect\citeauthoryear{Cooper et~al.}{2012}]{Cooper2012}
\begin{barticle}
\bauthor{\bsnm{Cooper}, \binits{L.}},
\bauthor{\bsnm{Walls}, \binits{R.L.}},
\bauthor{\bsnm{Elser}, \binits{J.}},
\bauthor{\bsnm{Gandolfo}, \binits{M.A.}},
\bauthor{\bsnm{Stevenson}, \binits{D.W.}},
\bauthor{\bsnm{Smith}, \binits{B.}},
\bauthor{\bsnm{Preece}, \binits{J.}},
\bauthor{\bsnm{Athreya}, \binits{B.}},
\bauthor{\bsnm{Mungall}, \binits{C.J.}},
\bauthor{\bsnm{Rensing}, \binits{S.}},
\bauthor{\bsnm{Hiss}, \binits{M.}},
\bauthor{\bsnm{Lang}, \binits{D.}},
\bauthor{\bsnm{Reski}, \binits{R.}},
\bauthor{\bsnm{Berardini}, \binits{T.Z.}},
\bauthor{\bsnm{Li}, \binits{D.}},
\bauthor{\bsnm{Huala}, \binits{E.}},
\bauthor{\bsnm{Schaeffer}, \binits{M.}},
\bauthor{\bsnm{Menda}, \binits{N.}},
\bauthor{\bsnm{Arnaud}, \binits{E.}},
\bauthor{\bsnm{Shrestha}, \binits{R.}},
\bauthor{\bsnm{Yamazaki}, \binits{Y.}},
\bauthor{\bsnm{Jaiswal}, \binits{P.}}:
\batitle{The plant ontology as a tool for comparative plant anatomy and genomic
  analyses}.
\bjtitle{Plant and Cell Physiology}
\bvolume{54}(\bissue{2}),
\bfpage{1}--\blpage{1}
(\byear{2012})
\doiurl{10.1093/pcp/pcs163}
\end{barticle}
\endbibitem

%%% 26
\bibitem[\protect\citeauthoryear{Cooper et~al.}{2017}]{Cooper2017}
\begin{barticle}
\bauthor{\bsnm{Cooper}, \binits{L.}},
\bauthor{\bsnm{Meier}, \binits{A.}},
\bauthor{\bsnm{Laporte}, \binits{M.-A.}},
\bauthor{\bsnm{Elser}, \binits{J.L.}},
\bauthor{\bsnm{Mungall}, \binits{C.}},
\bauthor{\bsnm{Sinn}, \binits{B.T.}},
\bauthor{\bsnm{Cavaliere}, \binits{D.}},
\bauthor{\bsnm{Carbon}, \binits{S.}},
\bauthor{\bsnm{Dunn}, \binits{N.A.}},
\bauthor{\bsnm{Smith}, \binits{B.}},
\bauthor{\bsnm{Qu}, \binits{B.}},
\bauthor{\bsnm{Preece}, \binits{J.}},
\bauthor{\bsnm{Zhang}, \binits{E.}},
\bauthor{\bsnm{Todorovic}, \binits{S.}},
\bauthor{\bsnm{Gkoutos}, \binits{G.}},
\bauthor{\bsnm{Doonan}, \binits{J.H.}},
\bauthor{\bsnm{Stevenson}, \binits{D.W.}},
\bauthor{\bsnm{Arnaud}, \binits{E.}},
\bauthor{\bsnm{Jaiswal}, \binits{P.}}:
\batitle{The planteome database: an integrated resource for reference
  ontologies, plant genomics and phenomics}.
\bjtitle{Nucleic Acids Research}
\bvolume{46}(\bissue{D1}),
\bfpage{1168}--\blpage{1180}
(\byear{2017})
\doiurl{10.1093/nar/gkx1152}
\end{barticle}
\endbibitem

%%% 27
\bibitem[\protect\citeauthoryear{Batchelor}{2020}]{Batchelor2020}
\begin{botherref}
\oauthor{\bsnm{Batchelor}, \binits{C.}}:
Chemical Reactions Ontology ({RXNO}).
The molecular process ontology {(MOP)} is distributed with the RXNO ontology
  for chemical reactions
(2020).
\url{https://github.com/rsc-ontologies/rxno}
\end{botherref}
\endbibitem

%%% 28
\bibitem[\protect\citeauthoryear{}{}]{MOP}
\begin{botherref}
\url{https://bioportal.bioontology.org/ontologies/MOP}
\end{botherref}
\endbibitem

%%% 29
\bibitem[\protect\citeauthoryear{Bandrowski et~al.}{2016}]{Bandrowski2016}
\begin{barticle}
\bauthor{\bsnm{Bandrowski}, \binits{A.}},
\bauthor{\bsnm{Brinkman}, \binits{R.}},
\bauthor{\bsnm{Brochhausen}, \binits{M.}},
\bauthor{\bsnm{Brush}, \binits{M.H.}},
\bauthor{\bsnm{Bug}, \binits{B.}},
\bauthor{\bsnm{Chibucos}, \binits{M.C.}},
\bauthor{\bsnm{Clancy}, \binits{K.}},
\bauthor{\bsnm{Courtot}, \binits{M.}},
\bauthor{\bsnm{Derom}, \binits{D.}},
\bauthor{\bsnm{Dumontier}, \binits{M.}},
\bauthor{\bsnm{Fan}, \binits{L.}},
\bauthor{\bsnm{Fostel}, \binits{J.}},
\bauthor{\bsnm{Fragoso}, \binits{G.}},
\bauthor{\bsnm{Gibson}, \binits{F.}},
\bauthor{\bsnm{Gonzalez-Beltran}, \binits{A.}},
\bauthor{\bsnm{Haendel}, \binits{M.A.}},
\bauthor{\bsnm{He}, \binits{Y.}},
\bauthor{\bsnm{Heiskanen}, \binits{M.}},
\bauthor{\bsnm{Hernandez-Boussard}, \binits{T.}},
\bauthor{\bsnm{Jensen}, \binits{M.}},
\bauthor{\bsnm{Lin}, \binits{Y.}},
\bauthor{\bsnm{Lister}, \binits{A.L.}},
\bauthor{\bsnm{Lord}, \binits{P.}},
\bauthor{\bsnm{Malone}, \binits{J.}},
\bauthor{\bsnm{Manduchi}, \binits{E.}},
\bauthor{\bsnm{McGee}, \binits{M.}},
\bauthor{\bsnm{Morrison}, \binits{N.}},
\bauthor{\bsnm{Overton}, \binits{J.A.}},
\bauthor{\bsnm{Parkinson}, \binits{H.}},
\bauthor{\bsnm{Peters}, \binits{B.}},
\bauthor{\bsnm{Rocca-Serra}, \binits{P.}},
\bauthor{\bsnm{Ruttenberg}, \binits{A.}},
\bauthor{\bsnm{Sansone}, \binits{S.-A.}},
\bauthor{\bsnm{Scheuermann}, \binits{R.H.}},
\bauthor{\bsnm{Schober}, \binits{D.}},
\bauthor{\bsnm{Smith}, \binits{B.}},
\bauthor{\bsnm{Soldatova}, \binits{L.N.}},
\bauthor{\bsnm{Stoeckert}, \binits{C.J.}},
\bauthor{\bsnm{Taylor}, \binits{C.F.}},
\bauthor{\bsnm{Torniai}, \binits{C.}},
\bauthor{\bsnm{Turner}, \binits{J.A.}},
\bauthor{\bsnm{Vita}, \binits{R.}},
\bauthor{\bsnm{Whetzel}, \binits{P.L.}},
\bauthor{\bsnm{Zheng}, \binits{J.}}:
\batitle{The ontology for biomedical investigations}.
\bjtitle{{PLOS} {ONE}}
\bvolume{11}(\bissue{4}),
\bfpage{0154556}
(\byear{2016})
\doiurl{10.1371/journal.pone.0154556}
\end{barticle}
\endbibitem

%%% 30
\bibitem[\protect\citeauthoryear{Gkoutos et~al.}{2017}]{Gkoutos2017}
\begin{barticle}
\bauthor{\bsnm{Gkoutos}, \binits{G.V.}},
\bauthor{\bsnm{Schofield}, \binits{P.N.}},
\bauthor{\bsnm{Hoehndorf}, \binits{R.}}:
\batitle{The anatomy of phenotype ontologies: principles, properties and
  applications}.
\bjtitle{Briefings in Bioinformatics}
\bvolume{19}(\bissue{5}),
\bfpage{1008}--\blpage{1021}
(\byear{2017})
\doiurl{10.1093/bib/bbx035}
\end{barticle}
\endbibitem

%%% 31
\bibitem[\protect\citeauthoryear{Gkoutos et~al.}{2004}]{Gkoutos2004}
\begin{barticle}
\bauthor{\bsnm{Gkoutos}, \binits{G.V.}},
\bauthor{\bsnm{Green}, \binits{E.C.}},
\bauthor{\bsnm{Mallon}, \binits{A.-M.}},
\bauthor{\bsnm{Hancock}, \binits{J.M.}},
\bauthor{\bsnm{Davidson}, \binits{D.}}:
\batitle{Using ontologies to describe mouse phenotypes}.
\bjtitle{Genome Biology}
\bvolume{6}(\bissue{1}),
\bfpage{8}
(\byear{2004})
\doiurl{10.1186/gb-2004-6-1-r8}
\end{barticle}
\endbibitem

%%% 32
\bibitem[\protect\citeauthoryear{}{}]{Ecocore}
\begin{botherref}
\url{https://github.com/EcologicalSemantics/ecocore}
\end{botherref}
\endbibitem

%%% 33
\bibitem[\protect\citeauthoryear{Campos et~al.}{2018}]{Campos2018}
\begin{bchapter}
\bauthor{\bsnm{Campos}, \binits{R.}},
\bauthor{\bsnm{Mangaravite}, \binits{V.}},
\bauthor{\bsnm{Pasquali}, \binits{A.}},
\bauthor{\bsnm{Jorge}, \binits{A.M.}},
\bauthor{\bsnm{Nunes}, \binits{C.}},
\bauthor{\bsnm{Jatowt}, \binits{A.}}:
\bctitle{A text feature based automatic keyword extraction method for single
  documents}.
In: \beditor{\bsnm{Pasi}, \binits{G.}},
\beditor{\bsnm{Piwowarski}, \binits{B.}},
\beditor{\bsnm{Azzopardi}, \binits{L.}},
\beditor{\bsnm{Hanbury}, \binits{A.}} (eds.)
\bbtitle{Advances in Information Retrieval},
pp. \bfpage{684}--\blpage{691}.
\bpublisher{Springer},
\blocation{Cham}
(\byear{2018})
\end{bchapter}
\endbibitem

%%% 34
\bibitem[\protect\citeauthoryear{King et~al.}{2017}]{King2017}
\begin{barticle}
\bauthor{\bsnm{King}, \binits{G.}},
\bauthor{\bsnm{Lam}, \binits{P.}},
\bauthor{\bsnm{Roberts}, \binits{M.E.}}:
\batitle{Computer-assisted keyword and document set discovery from unstructured
  text}.
\bjtitle{American Journal of Political Science}
\bvolume{61}(\bissue{4}),
\bfpage{971}--\blpage{988}
(\byear{2017})
\doiurl{10.1111/ajps.12291}
\end{barticle}
\endbibitem

%%% 35
\bibitem[\protect\citeauthoryear{Noy}{2001}]{noy2001}
\begin{bchapter}
\bauthor{\bsnm{Noy}, \binits{N.}}:
\bctitle{Ontology development 101: A guide to creating your first ontology}.
(\byear{2001}).
\burl{https://api.semanticscholar.org/CorpusID:500106}
\end{bchapter}
\endbibitem

%%% 36
\bibitem[\protect\citeauthoryear{Wu et~al.}{2012}]{wu2012}
\begin{bchapter}
\bauthor{\bsnm{Wu}, \binits{W.}},
\bauthor{\bsnm{Li}, \binits{H.}},
\bauthor{\bsnm{Wang}, \binits{H.}},
\bauthor{\bsnm{Zhu}, \binits{K.Q.}}:
\bctitle{Probase: A probabilistic taxonomy for text understanding}.
In: \bbtitle{Proceedings of the 2012 ACM SIGMOD International Conference on
  Management of Data}.
\bsertitle{SIGMOD '12},
pp. \bfpage{481}--\blpage{492}.
\bpublisher{Association for Computing Machinery},
\blocation{New York, NY, USA}
(\byear{2012}).
\doiurl{10.1145/2213836.2213891} .
\burl{https://doi.org/10.1145/2213836.2213891}
\end{bchapter}
\endbibitem

%%% 37
\bibitem[\protect\citeauthoryear{McDonald et~al.}{2016}]{mcdonald2016}
\begin{bchapter}
\bauthor{\bsnm{McDonald}, \binits{D.}},
\bauthor{\bsnm{Friedman}, \binits{S.}},
\bauthor{\bsnm{Paullada}, \binits{A.}},
\bauthor{\bsnm{Bobrow}, \binits{R.}},
\bauthor{\bsnm{Burstein}, \binits{M.}}:
\bctitle{Extending biology models with deep nlp over scientific articles}.
In: \bbtitle{AAAI Workshops}
(\byear{2016}).
\burl{https://www.aaai.org/ocs/index.php/WS/AAAIW16/paper/view/12615/12418}
\end{bchapter}
\endbibitem

%%% 38
\bibitem[\protect\citeauthoryear{Nasar et~al.}{2019}]{Nasar2019}
\begin{barticle}
\bauthor{\bsnm{Nasar}, \binits{Z.}},
\bauthor{\bsnm{Jaffry}, \binits{S.W.}},
\bauthor{\bsnm{Malik}, \binits{M.K.}}:
\batitle{Textual keyword extraction and summarization: State-of-the-art}.
\bjtitle{Information Processing and Management}
\bvolume{56}(\bissue{6}),
\bfpage{102088}
(\byear{2019})
\end{barticle}
\endbibitem

%%% 39
\bibitem[\protect\citeauthoryear{Hegde and Patil}{2020}]{Hegde2020}
\begin{botherref}
\oauthor{\bsnm{Hegde}, \binits{C.}},
\oauthor{\bsnm{Patil}, \binits{S.}}:
Unsupervised Paraphrase Generation using Pre-trained Language Models.
arXiv
(2020).
\doiurl{10.48550/ARXIV.2006.05477}
\end{botherref}
\endbibitem

%%% 40
\bibitem[\protect\citeauthoryear{Onan et~al.}{2016}]{Onan2016}
\begin{barticle}
\bauthor{\bsnm{Onan}, \binits{A.}},
\bauthor{\bsnm{Koruko{\u{g}}lu}, \binits{S.}},
\bauthor{\bsnm{Bulut}, \binits{H.}}:
\batitle{Ensemble of keyword extraction methods and classifiers in text
  classification}.
\bjtitle{Expert Systems with Applications}
\bvolume{57},
\bfpage{232}--\blpage{247}
(\byear{2016})
\doiurl{10.1016/j.eswa.2016.03.045}
\end{barticle}
\endbibitem

%%% 41
\bibitem[\protect\citeauthoryear{Rinartha and Kartika}{2021}]{Rinartha2021}
\begin{bchapter}
\bauthor{\bsnm{Rinartha}, \binits{K.}},
\bauthor{\bsnm{Kartika}, \binits{L.G.S.}}:
\bctitle{Rapid automatic keyword extraction and word frequency in scientific
  article keywords extraction}.
In: \bbtitle{2021 3rd International Conference on Cybernetics and Intelligent
  System ({ICORIS})}.
\bpublisher{{IEEE}}, \blocation{???}
(\byear{2021}).
\doiurl{10.1109/icoris52787.2021.9649458}
\end{bchapter}
\endbibitem

%%% 42
\bibitem[\protect\citeauthoryear{Papagiannopoulou}{2021}]{papagiannopoulou2021}
\begin{botherref}
\oauthor{\bsnm{Papagiannopoulou}, \binits{E.}}:
Keyphrase extraction techniques.
PhD thesis,
ARISTOTLE UNIVERSITY OF THESSALONIKI
(2021)
\end{botherref}
\endbibitem

%%% 43
\bibitem[\protect\citeauthoryear{Hasan and Ng}{2014}]{Hasan2014}
\begin{bchapter}
\bauthor{\bsnm{Hasan}, \binits{K.}},
\bauthor{\bsnm{Ng}, \binits{V.}}:
\bctitle{Automatic keyphrase extraction: A survey of the state of the art}.
In: \bbtitle{Proceedings of the 52nd Annual Meeting of the Association for
  Computational Linguistics (Volume 1: Long Papers)}.
\bpublisher{Association for Computational Linguistics}, \blocation{???}
(\byear{2014}).
\doiurl{10.3115/v1/p14-1119}
\end{bchapter}
\endbibitem

%%% 44
\bibitem[\protect\citeauthoryear{Juuti et~al.}{2020}]{Juuti2020}
\begin{bchapter}
\bauthor{\bsnm{Juuti}, \binits{M.}},
\bauthor{\bsnm{Gröndahl}, \binits{T.}},
\bauthor{\bsnm{Flanagan}, \binits{A.}},
\bauthor{\bsnm{Asokan}, \binits{N.}}:
\bctitle{A little goes a long way: Improving toxic language classification
  despite data scarcity}.
In: \bbtitle{Findings of the Association for Computational Linguistics: {EMNLP}
  2020}.
\bpublisher{Association for Computational Linguistics}, \blocation{???}
(\byear{2020}).
\doiurl{10.18653/v1/2020.findings-emnlp.269}
\end{bchapter}
\endbibitem

%%% 45
\bibitem[\protect\citeauthoryear{Dai et~al.}{2023}]{Dai2023}
\begin{botherref}
\oauthor{\bsnm{Dai}, \binits{H.}},
\oauthor{\bsnm{Liu}, \binits{Z.}},
\oauthor{\bsnm{Liao}, \binits{W.}},
\oauthor{\bsnm{Huang}, \binits{X.}},
\oauthor{\bsnm{Cao}, \binits{Y.}},
\oauthor{\bsnm{Wu}, \binits{Z.}},
\oauthor{\bsnm{Zhao}, \binits{L.}},
\oauthor{\bsnm{Xu}, \binits{S.}},
\oauthor{\bsnm{Liu}, \binits{W.}},
\oauthor{\bsnm{Liu}, \binits{N.}},
\oauthor{\bsnm{Li}, \binits{S.}},
\oauthor{\bsnm{Zhu}, \binits{D.}},
\oauthor{\bsnm{Cai}, \binits{H.}},
\oauthor{\bsnm{Sun}, \binits{L.}},
\oauthor{\bsnm{Li}, \binits{Q.}},
\oauthor{\bsnm{Shen}, \binits{D.}},
\oauthor{\bsnm{Liu}, \binits{T.}},
\oauthor{\bsnm{Li}, \binits{X.}}:
AugGPT: Leveraging ChatGPT for Text Data Augmentation.
arXiv
(2023).
\doiurl{10.48550/ARXIV.2302.13007}
\end{botherref}
\endbibitem

%%% 46
\bibitem[\protect\citeauthoryear{Yoo et~al.}{2021}]{Yoo2021}
\begin{botherref}
\oauthor{\bsnm{Yoo}, \binits{K.M.}},
\oauthor{\bsnm{Park}, \binits{D.}},
\oauthor{\bsnm{Kang}, \binits{J.}},
\oauthor{\bsnm{Lee}, \binits{S.-W.}},
\oauthor{\bsnm{Park}, \binits{W.}}:
GPT3Mix: Leveraging Large-scale Language Models for Text Augmentation.
arXiv
(2021).
\doiurl{10.48550/ARXIV.2104.08826}
\end{botherref}
\endbibitem

%%% 47
\bibitem[\protect\citeauthoryear{Anaby-Tavor et~al.}{2019}]{AnabyTavor2019}
\begin{botherref}
\oauthor{\bsnm{Anaby-Tavor}, \binits{A.}},
\oauthor{\bsnm{Carmeli}, \binits{B.}},
\oauthor{\bsnm{Goldbraich}, \binits{E.}},
\oauthor{\bsnm{Kantor}, \binits{A.}},
\oauthor{\bsnm{Kour}, \binits{G.}},
\oauthor{\bsnm{Shlomov}, \binits{S.}},
\oauthor{\bsnm{Tepper}, \binits{N.}},
\oauthor{\bsnm{Zwerdling}, \binits{N.}}:
Not Enough Data? Deep Learning to the Rescue!
arXiv
(2019).
\doiurl{10.48550/ARXIV.1911.03118}
\end{botherref}
\endbibitem

%%% 48
\bibitem[\protect\citeauthoryear{Radford et~al.}{2019}]{radford2019}
\begin{barticle}
\bauthor{\bsnm{Radford}, \binits{A.}},
\bauthor{\bsnm{Wu}, \binits{J.}},
\bauthor{\bsnm{Child}, \binits{R.}},
\bauthor{\bsnm{Luan}, \binits{D.}},
\bauthor{\bsnm{Amodei}, \binits{D.}},
\bauthor{\bsnm{Sutskever}, \binits{I.}}, \betal:
\batitle{Language models are unsupervised multitask learners}.
\bjtitle{OpenAI blog}
\bvolume{1}(\bissue{8}),
\bfpage{9}
(\byear{2019})
\end{barticle}
\endbibitem

%%% 49
\bibitem[\protect\citeauthoryear{Brown et~al.}{2020}]{Brown2020}
\begin{botherref}
\oauthor{\bsnm{Brown}, \binits{T.B.}},
\oauthor{\bsnm{Mann}, \binits{B.}},
\oauthor{\bsnm{Ryder}, \binits{N.}},
\oauthor{\bsnm{Subbiah}, \binits{M.}},
\oauthor{\bsnm{Kaplan}, \binits{J.}},
\oauthor{\bsnm{Dhariwal}, \binits{P.}},
\oauthor{\bsnm{Neelakantan}, \binits{A.}},
\oauthor{\bsnm{Shyam}, \binits{P.}},
\oauthor{\bsnm{Sastry}, \binits{G.}},
\oauthor{\bsnm{Askell}, \binits{A.}},
\oauthor{\bsnm{Agarwal}, \binits{S.}},
\oauthor{\bsnm{Herbert-Voss}, \binits{A.}},
\oauthor{\bsnm{Krueger}, \binits{G.}},
\oauthor{\bsnm{Henighan}, \binits{T.}},
\oauthor{\bsnm{Child}, \binits{R.}},
\oauthor{\bsnm{Ramesh}, \binits{A.}},
\oauthor{\bsnm{Ziegler}, \binits{D.M.}},
\oauthor{\bsnm{Wu}, \binits{J.}},
\oauthor{\bsnm{Winter}, \binits{C.}},
\oauthor{\bsnm{Hesse}, \binits{C.}},
\oauthor{\bsnm{Chen}, \binits{M.}},
\oauthor{\bsnm{Sigler}, \binits{E.}},
\oauthor{\bsnm{Litwin}, \binits{M.}},
\oauthor{\bsnm{Gray}, \binits{S.}},
\oauthor{\bsnm{Chess}, \binits{B.}},
\oauthor{\bsnm{Clark}, \binits{J.}},
\oauthor{\bsnm{Berner}, \binits{C.}},
\oauthor{\bsnm{McCandlish}, \binits{S.}},
\oauthor{\bsnm{Radford}, \binits{A.}},
\oauthor{\bsnm{Sutskever}, \binits{I.}},
\oauthor{\bsnm{Amodei}, \binits{D.}}:
Language Models are Few-Shot Learners.
arXiv
(2020).
\doiurl{10.48550/ARXIV.2005.14165}
\end{botherref}
\endbibitem

%%% 50
\bibitem[\protect\citeauthoryear{Wang et~al.}{2021}]{Wang2021}
\begin{botherref}
\oauthor{\bsnm{Wang}, \binits{Z.}},
\oauthor{\bsnm{Yu}, \binits{A.W.}},
\oauthor{\bsnm{Firat}, \binits{O.}},
\oauthor{\bsnm{Cao}, \binits{Y.}}:
Towards Zero-Label Language Learning.
arXiv
(2021).
\doiurl{10.48550/ARXIV.2109.09193}
\end{botherref}
\endbibitem

%%% 51
\bibitem[\protect\citeauthoryear{{OpenAI}}{2023}]{OpenAI2023}
\begin{botherref}
\oauthor{\bsnm{{OpenAI}}}:
GPT-4 Technical Report.
arXiv
(2023).
\doiurl{10.48550/ARXIV.2303.08774}
\end{botherref}
\endbibitem

%%% 52
\bibitem[\protect\citeauthoryear{Touvron et~al.}{2023a}]{Touvron2023}
\begin{botherref}
\oauthor{\bsnm{Touvron}, \binits{H.}},
\oauthor{\bsnm{Lavril}, \binits{T.}},
\oauthor{\bsnm{Izacard}, \binits{G.}},
\oauthor{\bsnm{Martinet}, \binits{X.}},
\oauthor{\bsnm{Lachaux}, \binits{M.-A.}},
\oauthor{\bsnm{Lacroix}, \binits{T.}},
\oauthor{\bsnm{Rozière}, \binits{B.}},
\oauthor{\bsnm{Goyal}, \binits{N.}},
\oauthor{\bsnm{Hambro}, \binits{E.}},
\oauthor{\bsnm{Azhar}, \binits{F.}},
\oauthor{\bsnm{Rodriguez}, \binits{A.}},
\oauthor{\bsnm{Joulin}, \binits{A.}},
\oauthor{\bsnm{Grave}, \binits{E.}},
\oauthor{\bsnm{Lample}, \binits{G.}}:
LLaMA: Open and Efficient Foundation Language Models.
arXiv
(2023).
\doiurl{10.48550/ARXIV.2302.13971}
\end{botherref}
\endbibitem

%%% 53
\bibitem[\protect\citeauthoryear{Touvron et~al.}{2023b}]{Touvron2023a}
\begin{botherref}
\oauthor{\bsnm{Touvron}, \binits{H.}},
\oauthor{\bsnm{Martin}, \binits{L.}},
\oauthor{\bsnm{Stone}, \binits{K.}},
\oauthor{\bsnm{Albert}, \binits{P.}},
\oauthor{\bsnm{Almahairi}, \binits{A.}},
\oauthor{\bsnm{Babaei}, \binits{Y.}},
\oauthor{\bsnm{Bashlykov}, \binits{N.}},
\oauthor{\bsnm{Batra}, \binits{S.}},
\oauthor{\bsnm{Bhargava}, \binits{P.}},
\oauthor{\bsnm{Bhosale}, \binits{S.}},
\oauthor{\bsnm{Bikel}, \binits{D.}},
\oauthor{\bsnm{Blecher}, \binits{L.}},
\oauthor{\bsnm{Ferrer}, \binits{C.C.}},
\oauthor{\bsnm{Chen}, \binits{M.}},
\oauthor{\bsnm{Cucurull}, \binits{G.}},
\oauthor{\bsnm{Esiobu}, \binits{D.}},
\oauthor{\bsnm{Fernandes}, \binits{J.}},
\oauthor{\bsnm{Fu}, \binits{J.}},
\oauthor{\bsnm{Fu}, \binits{W.}},
\oauthor{\bsnm{Fuller}, \binits{B.}},
\oauthor{\bsnm{Gao}, \binits{C.}},
\oauthor{\bsnm{Goswami}, \binits{V.}},
\oauthor{\bsnm{Goyal}, \binits{N.}},
\oauthor{\bsnm{Hartshorn}, \binits{A.}},
\oauthor{\bsnm{Hosseini}, \binits{S.}},
\oauthor{\bsnm{Hou}, \binits{R.}},
\oauthor{\bsnm{Inan}, \binits{H.}},
\oauthor{\bsnm{Kardas}, \binits{M.}},
\oauthor{\bsnm{Kerkez}, \binits{V.}},
\oauthor{\bsnm{Khabsa}, \binits{M.}},
\oauthor{\bsnm{Kloumann}, \binits{I.}},
\oauthor{\bsnm{Korenev}, \binits{A.}},
\oauthor{\bsnm{Koura}, \binits{P.S.}},
\oauthor{\bsnm{Lachaux}, \binits{M.-A.}},
\oauthor{\bsnm{Lavril}, \binits{T.}},
\oauthor{\bsnm{Lee}, \binits{J.}},
\oauthor{\bsnm{Liskovich}, \binits{D.}},
\oauthor{\bsnm{Lu}, \binits{Y.}},
\oauthor{\bsnm{Mao}, \binits{Y.}},
\oauthor{\bsnm{Martinet}, \binits{X.}},
\oauthor{\bsnm{Mihaylov}, \binits{T.}},
\oauthor{\bsnm{Mishra}, \binits{P.}},
\oauthor{\bsnm{Molybog}, \binits{I.}},
\oauthor{\bsnm{Nie}, \binits{Y.}},
\oauthor{\bsnm{Poulton}, \binits{A.}},
\oauthor{\bsnm{Reizenstein}, \binits{J.}},
\oauthor{\bsnm{Rungta}, \binits{R.}},
\oauthor{\bsnm{Saladi}, \binits{K.}},
\oauthor{\bsnm{Schelten}, \binits{A.}},
\oauthor{\bsnm{Silva}, \binits{R.}},
\oauthor{\bsnm{Smith}, \binits{E.M.}},
\oauthor{\bsnm{Subramanian}, \binits{R.}},
\oauthor{\bsnm{Tan}, \binits{X.E.}},
\oauthor{\bsnm{Tang}, \binits{B.}},
\oauthor{\bsnm{Taylor}, \binits{R.}},
\oauthor{\bsnm{Williams}, \binits{A.}},
\oauthor{\bsnm{Kuan}, \binits{J.X.}},
\oauthor{\bsnm{Xu}, \binits{P.}},
\oauthor{\bsnm{Yan}, \binits{Z.}},
\oauthor{\bsnm{Zarov}, \binits{I.}},
\oauthor{\bsnm{Zhang}, \binits{Y.}},
\oauthor{\bsnm{Fan}, \binits{A.}},
\oauthor{\bsnm{Kambadur}, \binits{M.}},
\oauthor{\bsnm{Narang}, \binits{S.}},
\oauthor{\bsnm{Rodriguez}, \binits{A.}},
\oauthor{\bsnm{Stojnic}, \binits{R.}},
\oauthor{\bsnm{Edunov}, \binits{S.}},
\oauthor{\bsnm{Scialom}, \binits{T.}}:
Llama 2: Open Foundation and Fine-Tuned Chat Models.
arXiv
(2023).
\doiurl{10.48550/ARXIV.2307.09288}
\end{botherref}
\endbibitem

%%% 54
\bibitem[\protect\citeauthoryear{Perez and Wang}{2017}]{Perez2017}
\begin{botherref}
\oauthor{\bsnm{Perez}, \binits{L.}},
\oauthor{\bsnm{Wang}, \binits{J.}}:
The Effectiveness of Data Augmentation in Image Classification using Deep
  Learning.
arXiv
(2017).
\doiurl{10.48550/ARXIV.1712.04621}
\end{botherref}
\endbibitem

%%% 55
\bibitem[\protect\citeauthoryear{Jin et~al.}{2023}]{jin2023genegpt}
\begin{botherref}
\oauthor{\bsnm{Jin}, \binits{Q.}},
\oauthor{\bsnm{Yang}, \binits{Y.}},
\oauthor{\bsnm{Chen}, \binits{Q.}},
\oauthor{\bsnm{Lu}, \binits{Z.}}:
GeneGPT: Augmenting Large Language Models with Domain Tools for Improved Access
  to Biomedical Information
(2023)
\end{botherref}
\endbibitem

\end{thebibliography}

\end{document}